\documentclass[acmsmall,screen,nonacm]{acmart}
\renewcommand\footnotetextcopyrightpermission[1]{}
\def\@copyrightspace{\relax}
\makeatletter
\renewcommand\@formatdoi[1]{\ignorespaces}
\makeatother

\AtBeginDocument{%
  \providecommand\BibTeX{{%
    \normalfont B\kern-0.5em{\scshape i\kern-0.25em b}\kern-0.8em\TeX}}}

\usepackage[euler]{textgreek}
\usepackage{amsmath}
\def\SPSB#1#2{\rlap{\textsuperscript{#1}}\SB{#2}}
\def\SB#1{\textsubscript{#1}}
\newcommand\w{0.48}
\newcommand\h{0.45}
\usepackage{subfig}
\usepackage{multirow}
\usepackage{multicol}
\usepackage{csvsimple}
\usepackage{longtable}
\usepackage{booktabs}
\usepackage{float}
\usepackage{varwidth}
\usepackage{blindtext}
\usepackage{nameref}
\usepackage{pdfpages}

\setcopyright{none}

\graphicspath{{Prediction-Images/}}

\begin{document}

\title{The Past, Present, and Future of COVID-19: A Data-Driven Perspective}

\author{Ajwad Akil}
\email{stranger.things.warp@gmail.com}
\affiliation{%
  \institution{Bangladesh University of Engineering and Technology}
}
\author{Ishrat Jahan Eliza}
\email{elizaan199834@gmail.com}
\affiliation{%
  \institution{Bangladesh University of Engineering and Technology}
}

\author{Md. Hasibul Hussain Hisham}
\affiliation{%
  \institution{Bangladesh University of Engineering and Technology}
}
\email{mdhasibulhussainh@gmail.com}

\author{Fahim Morshed}
\affiliation{%
  \institution{Bangladesh University of Engineering and Technology}
}
\email{1605077@ugrad.cse.buet.ac.bd>}

\author{Nazmus Sakib}
\affiliation{%
  \institution{Bangladesh University of Engineering and Technology}
}
\email{nsakib280@gmail.com}

\author{Nuwaisir Rabi}
\affiliation{%
  \institution{Bangladesh University of Engineering and Technology}
}
\email{nuwaisir1998@gmail.com}

\author{Abir Mohammad Turza}
\affiliation{%
  \institution{Bangladesh University of Engineering and Technology}
}
\email{abir1turza@gmail.com}

\author{Sriram Chellappan}
\affiliation{%
  \institution{University of South Florida}
}
\email{sriramc@usf.edu}

\author{A. B. M. Alim Al Islam}
\affiliation{%
  \institution{Bangladesh University of Engineering and Technology}
}
\email{razi_bd@yahoo.com}







\begin{abstract}
Epidemics and pandemics have ravaged human life since time. To combat these, novel ideas have always been created and deployed by humanity, with varying degrees of success. At this very moment, the COVID-19 pandemic is the singular global health crisis. Now, perhaps for the first time in human history, almost the whole of humanity is experiencing some form of hardship as a result of one invisible pathogen.  This once again entails novel ideas for quick eradication, healing and recovery, whether it is healthcare, banking, travel, education or any other. For efficient policy-making, clear trends of past, present and future are vital for policy-makers. With the global impacts of COVID-19 so severe, equally important is the analysis of correlations between disease spread and various socio-economic and environmental factors. Furthermore, all of these need to be presented in an integrated manner in real-time to facilitate efficient policy making. To address these issues, in this study, we report results on our development and deployment of a web-based integrated real-time operational dashboard as an important decision support system for COVID-19. In our study, we conducted data-driven analysis based on available data from diverse authenticated sources to predict  upcoming consequences of the pandemic through rigorous modeling and statistical analyses. We also explored correlations between pandemic spread and important socio-economic and environmental factors. Furthermore, we also present how outcomes of our work can facilitate efficient policy making in this critical hour.
\end{abstract}



\maketitle

\section{Introduction}
The COVID-19 pandemic is a unique and un-precedent challenge that humanity is facing today. Never before in our history have we witnessed situations where social lockdowns are imposed across the world, travel is severely restricted, children and teens cannot go to study, and work from home is the new normal. There is no real positive news still about any cure or vaccine, and infection rates are massively increasing \cite{deathrate}.
The impact of the pandemic is near universal covering the arenas of politics \cite{political}, economics \cite{socioEco}, social \cite{socioEco}, psychological unrest \cite{psycho}, and mass panic \cite{panic} across the world. Mitigating these concerns is a severe challenge for leaders and policy-makers, and necessitates novel ideas that can advantage of technological progresses available now.

\begin{figure}[!h]
    \centering
    \includegraphics[width=0.9\textwidth, height = 0.55\textheight]{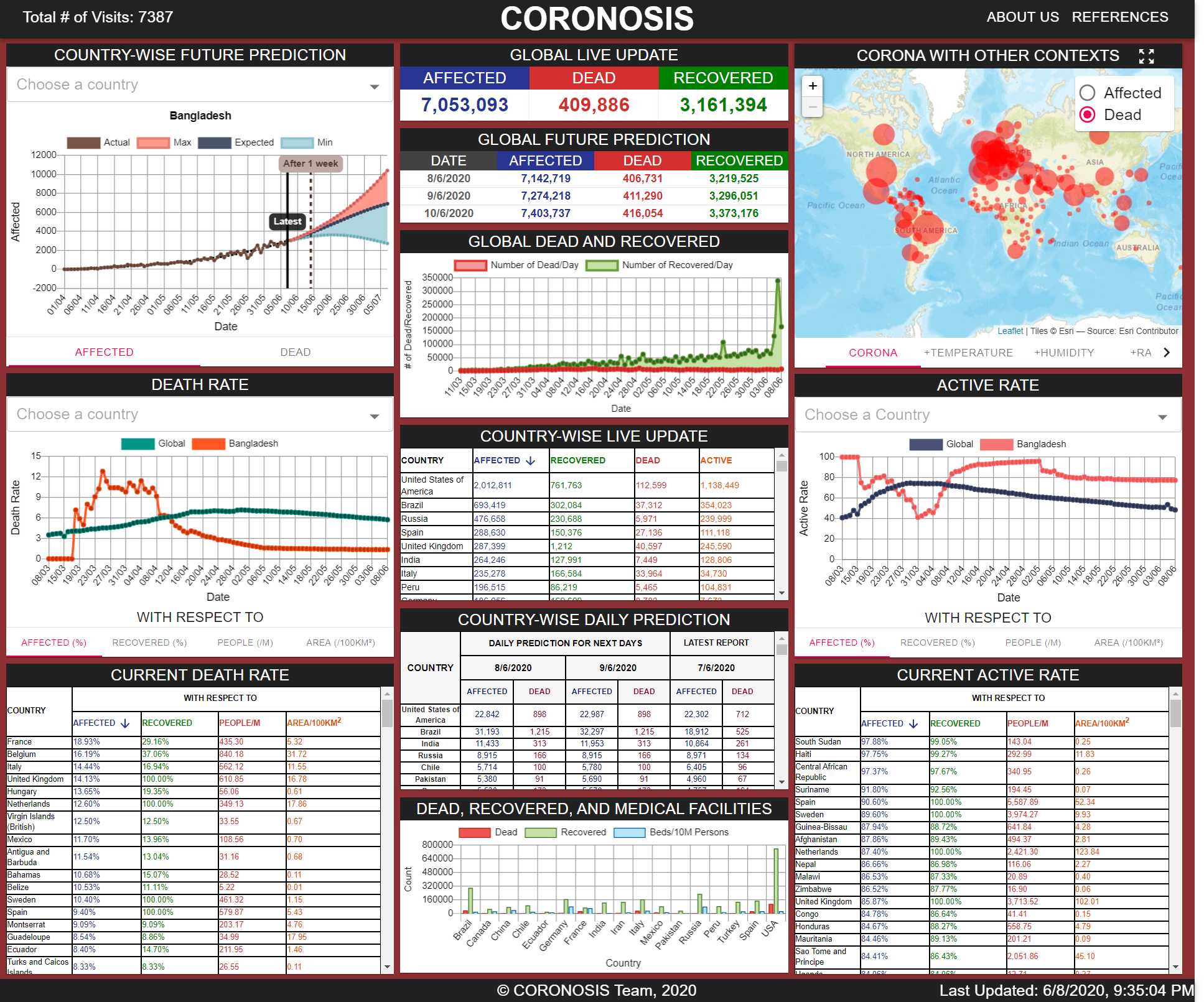}
    \caption{\textbf{CORONOSIS}}
    \label{fig:CORONOSIS}
\end{figure}

Towards addressing these challenges, and to gauge impact of any policies, trend analysis is vital. Towards facilitating this, our team  designed, developed and deployed a web-based integrated operational dashboard towards accomplishing two goals.
The first goal is the ability to compare real-time trends of pandemic spread across the globe. The second is predictive modeling of pandemic spread over time taking into account important socio-demographic and environmental factors using robust models (Figure: \ref{fig:CORONOSIS}). Accomplishing these goals can particularly provide useful insights and inferences for  policy makers. Therefore, we analyzed at length how  insights and inferences gleaned from our system can influence policy formulations directly or indirectly in diverse contexts to combat COVID-19.

\subsection{Challenges during COVID-19 Pandemic and The Need for Effective Data Analysis}

COVID-19 has so far affected millions of lives, and thousands have actually dead. Since the disease is highly contagious by nature (spreading through the air or touch), people are obligated to maintain appropriate social distancing and are advised to avoid mobility. This means children and young adults have to stay at home, and avoid going to schools/ colleges; daily wage laborers have completely lost their income; almost total shutdown of trains, planes and ships; gross inadequacies in the global supply chain (including food), severe stress on medical professionals and more. 

The global economy is severely suffering as a result. For example, according to the report~\cite{unidoreport} of UNIDO (United Nations Industrial Development Organization), the global GDP growth projections for 2020 vary between -8.8\% (by WTO)\cite{wtoreport} to 1\% (by UNDESA)\cite{undesareport}. Among all the projections of the GDP, the International Monetary Fund’s (IMF) is a well-accepted reference point for analyzing  economic impact of COVID-19. According to its projection,  global growth will shrink by -4.2\% directly due to the pandemic\cite{imfreport}, and global trade and investment will also take a massive hit. World trade volumes are anticipated to drop between  -32\% (by WTO)\cite{wtoreport} to -9\%(by IMF)\cite{imfreport} in 2020.   \newline

With no cures or vaccines in sights, it is incumbent upon all stakeholders - world leaders, policy makers, public health workers, researchers, economic advisors, and more importantly, common citizens to work together in combating this pandemic, and getting life back to normal. For this, among many things, information is paramount. Furthermore, the ability to rely on authentic sources, visualize information in a simple, and yet effective manner, compare global trends across space and time, and leverage state of the art robust analytical tools are very critical in this fight. We aim to address all these issues in this study.


\subsection{Motivation behind This Work And Our Research Context}
There exist a myriad of different web applications, news portals, and web tools that display  current situation of  COVID-19, albeit in a traditional way. Most  websites \cite{johnhopkins,worldometer} and news portals display  current statistics, maps, and charts showing  extent of affected, dead, and recovered along with  other general statistical data such as  death rates. Besides, only a handful of websites display  correlations, predictions, and analyses \cite{ihme}, also in a limited manner. Though these sources deliver  information, they lack  profound and insightful analyses and also lack in presenting diversified correlations between  pandemic spread and environmental as well as socio-economic factors in different parts of the world. Such diverse correlation results can help in significantly valuable information for policy makers as well as other stakeholders.


In terms of challenges, among multiple data sources and formats, there are issues related to - 1) bridging data sources together in real-time, 2) generating  other relevant information (socio-demographic, environmental etc.) most critical for policy makers, and 3) presenting them in an integrated manner. Such a system does not exist yet. This gap motivated us to create a fully functional and integrated dashboard, which we expect would be a one-stop source for significant useful data in combating COVID-19. The dashboard enables comparison of critical outcomes across countries in space and time, projects correlations among pandemic spread and important socio-demographic and economic factors, and also projects future trends based on robust models.


\subsection{Research Questions}

Being motivated by the research context, we focus on the following set of research questions.

\textit{RQ1}.\hspace{0.2cm}\emph{Is there any correlation (positive or negative) between the number of affected or dead due to the pandemic and different socio-economic as well as environmental contexts such as temperature, humidity, GDP, population density, literacy, etc.?}
    
    
\textit{RQ2}.\hspace{0.2cm}\emph{Can we predict the future outcomes (the number of affected or dead) of the pandemic with the existing data? If so, then how far we can go through spatiotemporal analyses?}
    
    
\textit{RQ3}.\hspace{0.2cm}\emph{How can we present outcomes of the analyses on correlations and future predictions in an integrated manner such that it can present a one-stop source of the related data?}

\textit{RQ4}.\hspace{0.2cm}\emph{How does data analysis based on correlation and future prediction as well as presenting them in an integrated manner affect formulating sustainable and robust policy making across the world?}

\subsection{Our Research Contributions}
To answer the above research questions, we make the following set of contributions in this paper.
\begin{itemize}
\item
We developed a new web-based integrated real-time operational dashboard as a one-stop decision support system through intelligent maps, graph-charts, info-graphs, and data analytics. 
We tailored the web application to enable visualization of data for the current pandemic. However, it can be extended for any epidemic or pandemic to come. 
\item
Through rigorous modeling and data analysis, we predicted future effects (the number of affected, dead, active, or recovered) of the pandemic for up to one month ahead. 
\item
Further, through in-depth statistical analyses, we explored correlation between the effect of the pandemic and a substantial number of socio-economic and environmental factors such as temperature, humidity, rainfall, GDP, population density, literacy, etc. We also considered medical facilities such as the number of hospital beds in our analysis.
\item
Finally, we analyzed how the outcomes of our work can facilitate efficient policy making. To do so, we present impacts of different results obtained through our analyses in different types of policy making.
\end{itemize}

\section{Related work}
We now situate our research in a body of related work examining how procuring pandemic information, data visualization, and data-driven mathematical analysis and modeling are used in a pandemic. 

\subsection{Pandemic Information and Visualization}
Transfiguring raw data into information and visualizing them in an interactive and intuitive way has been performed for other epidemics \cite{visua1}\cite{visua4} and pandemics \cite{visua2}\cite{visua3}. However, there also exists some other dashboards for COVID-19 pandemic \cite{johnhopkins}\cite{worldometer}\cite{visua5}. Like ours, all these dashboards provide intuitive information about the ongoing COVID-19 pandemic in an intuitive manner and contain in-depth analysis.

\subsection{Data-driven Mathematical Analysis}
Atmospheric conditions and socio-economic scenario of a region have always played a significant role in a virus outbreak. Such have already examined in previous epidemics \cite{papana1}and pandemics \cite{papana2}. Researchers all around the world performed analysis to understand an infectious disease in more robust way. They explored while correlating the pandemic with numerous aspects \cite{papana3}\cite{papana4}. In this ongoing COVID-19 pandemic, researchers also studied the pandemic by correlating the affect and cause with different socio-economical and geographic characteristics \cite{papana5}\cite{papana6}\cite{papana7}.

\subsection{Mathematical Prediction Model}
One essential step of mitigating the damages of a pandemic is to predict the pandemic in advance. Researchers have always been working off and on the time of a pandemic to let appropriate authorities tackle the upcoming difficulties. Since epidemics and pandemics impose a dreadful threat to the human civilization, researchers have tried to predict the upcoming pandemics \cite{predic1}\cite{predic2}. Besides, when in the time of an ongoing epidemics and pandemics, researchers tried predicted the future implications of the outbreak. This includes Influenza \cite{predic3}\cite{predic4}, Ebola \cite{predic5}, Swine Flu \cite{predic6} etc. 

Numerous prediction studies have also been carried out for the currently occurring COVID-19 pandemic. Various methods have been implemented while predicting the pandemic to facilitate the policy makers. There were map based approaches where mobility and migration was considered \cite{predic7}\cite{predic8}. There were different machine learning based approaches where basic machine learning models \cite{predic11} and image processing \cite{predic9}\cite{predic10} were used. There were mathematical models as well \cite{predic12}\cite{predic13}.   

\section{Methodology}
Methodology is defined as the systematic and theoretical analysis of various methods applied to a particular field of study. In terms of our research context, it encompasses the process of collecting our required data, how we designed our platform, what are the relevant technologies, what methods of analysis and analytics we performed and what challenges and shortcomings we faced along the way and how we resolved such issues.

\subsection{Development Tools}
In order to collect, analyze and build a reliable web based platform, we used number of frameworks, libraries of multiple languages coherently. The technologies we used are summarized below for better clarity and understanding of our platform. 
\subsubsection{Data Collection}
            We used various python based libraries/modules for the collection of data from a number of sources. We specifically choose to use these libraries because they provide simple methods to fetch, collect and clean data easily. Requests, json, csv are some of the prominent ones used for collecting data.
\subsubsection{Analysis}
            Data analysis is the core component of our research and to achieve this with ease, we relied heavily on the use of some python libraries related to data cleaning, analysis, visualization, statistics and prediction. Python provides easy to use and off-the shelf methods of these libraries that seemed suitable for our experiments. The libraries we used are: Numpy, Pandas, Matplotlib, Scikitlearn, and Plotly.
\subsubsection{Back End}
        Not only our web based platform, but also for the purpose of obtaining structured data for our research, we needed to make sure we have a reliable back end service that we can use seamlessly without interfering the web platform that it was primarily serving. Our framework of choice was the javascript based framework Express.js, which in turn is a framework of the run time node.js. It is known for speed and the ability to handle large amount of concurrent requests and scalability in the industries. Node.js provides a lot of dependencies that we used to achieve tasks at different parts of the server side. Some of the dependencies are(but not limited to): node-fetch, pg, axios, body-parser, concurrently, node-cache, etc.
\subsubsection{Front End}
        The front end potion of our web application provided a beautiful and interactive user interface consisting of plots, charts, maps and informative tables. The main library used to implement all of these was the popular javascript library React.js,which provides a structured way to generate front end part of the application through a component based system. Some of the dependencies and modules that was used(but not limited to) along with this to achieve various useful functionality are: Material-UI, React-leaflet, React-leaflet-control, D3-Scale, React-chartjs-3, Regression-js, etc.

\subsection{Development Architecture}
We developed a functional dashboard as a one-stop decision support system for providing important information and data visualization about the current ongoing pandemic COVID-19. The intent was to display aggregated information and data in an intuitive way through beautiful visualizations such as charts, graphs, info-graphs, interactive tables, intelligent and interactive maps, etc. \\
The application was made responsive with different screen sizes and resolutions.\\
The whole development architecture of the web application in accordance with the characteristics of the development tool can be divided into four specific categories - 1) Data Collection, 2) Analysis, 3) Back End, and 4)Front End.\\
We present an overview on each of the individual components and their relationships below.

\begin{figure}
    \centering
    \def\svgwidth{\columnwidth}
    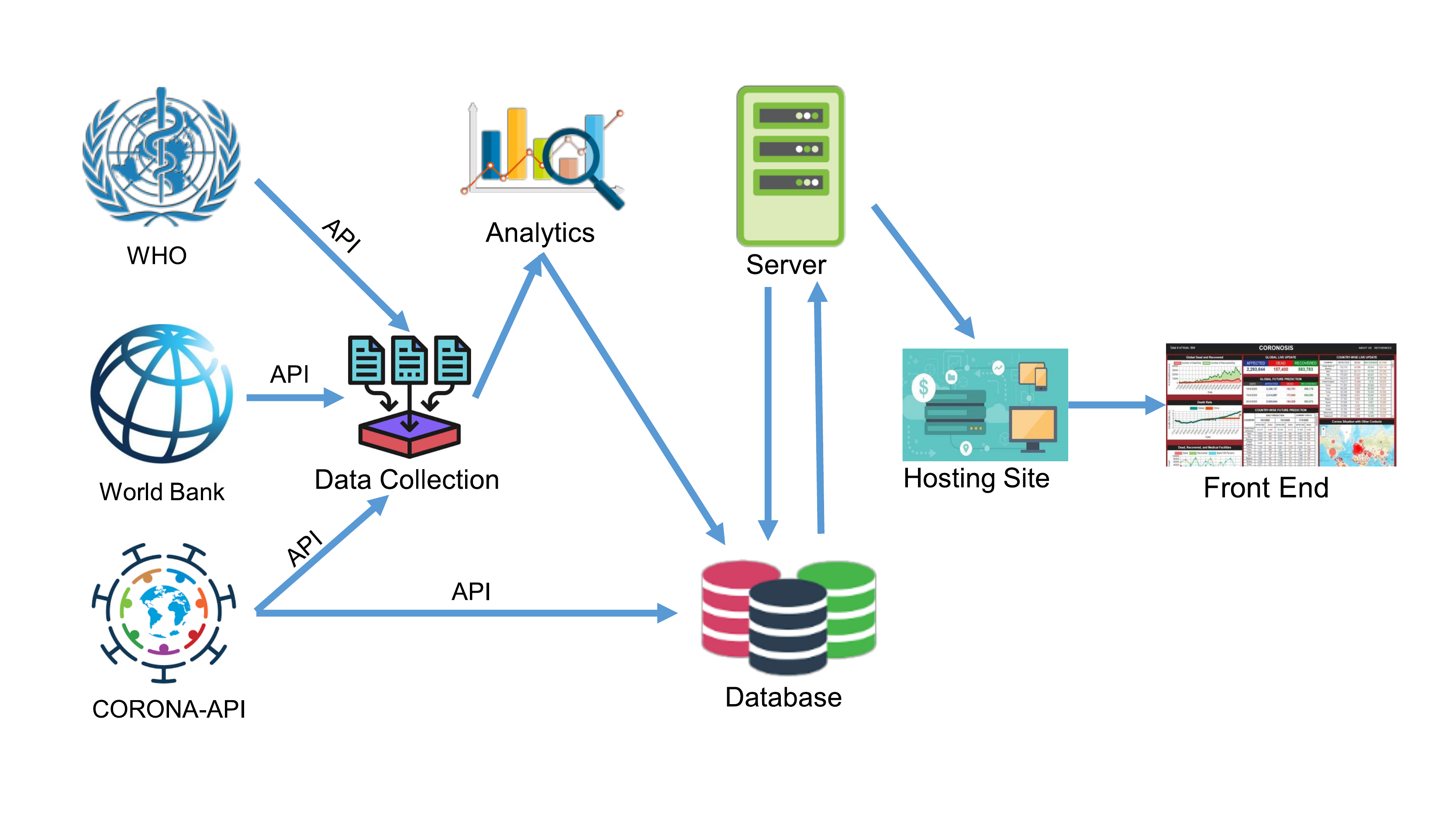
    \caption{\textbf{Development Architecture}}
\end{figure} 
    
\subsubsection{Data Collection}
    We collected our data from various reliable sources. We collected global live update from Covid19. Mathdro.Id \cite{mathidro}, data for Global Live Prediction from Corona-API \cite{covid-19-API}, data needed for Country-wise Future Prediction, Country-wise Daily Prediction from COVID-19 virus Tracker API \cite{coronatracker}, Dead, Recovered and Medical Facilities (BEDS/10M) from WHO \cite{worldbank}, Current Death Rate, Current Active Rate from NovelCOVID API \cite{death}, Humidity from Kaggle \cite{humKaggle}, Temperature, Rainfall, GDP, Literacy from WorldBank \cite{worldbank}, Area, Population, Population Density from REST Countries API \cite{population}, all other data from COVID-19 REST API \cite{covid-19-API}. After getting the data in csv format, we used Microsoft Excel and Python programming language to perform pre-processing. 
    It is noted that, as far as temperature and humidity data are concerned, we got country specific data from 22nd January to 1st April. So, we took the arithmetic mean of the temperature and humidity and used that for further statistical analysis. 
\subsubsection{Analysis}
    We retrieved data from APIs \cite{coronatracker,corona-api}. To use an API, we made a request to the remote web server, and retrieved the necessary data we needed. Our program fetched data using python's requests module and we received data in JSON format. After performing some preliminary data processing, we ended up with clean and ready to be analyzed datasets. Besides, the source's data came in the form of cumulative time-series. Thus, for some analyses, the time-series are transformed to represent the new cases of each day. More details about data analysis and analytic are mentioned in section 3.5 and 3.6.
    After doing some statistical as well as mathematical analysis and prediction based on analytics we passed our data to the back end team by pushing into the database so that they can process them to visualize onto the front end.
    
\subsubsection{Back End}
    The bulk of the server side development was to build a suitable API so that the front end part of the application can easily fetch the data as per required and also the end points can be used to fetch data for further research purposes.\\
    At the primary level, some important data such as GDP, literacy rate, average annual temperature, average annual rainfall, average humidity of the countries, population, country codes, country area were filtered and pushed to the database so that they can be accessed on demand. Some important data such as data for prediction are semi automatically pushed to the database after a few days. Besides, we also populated our database daily with the daily statistical summary of COVID-19 of all the countries of the world  with the help of an automated script.\\
    The server side was designed to have two specific types of API endpoints or URL routes. These URL routes, when hit upon correctly, provide JSON response in a suitable format. The data that we obtained from external API calls or database query could be majorly divided into three broad categories: (1) Constantly Changing Data, (2) Seldom Changing Data and (3) Almost Constant or Fixed Data. \newline
    We designed two types of URL routes or API endpoints to handle the different types of data. One of the endpoints was strictly used for handling the constantly changing data. When used with the rest of the URL, the response was that of live or constantly changing data (death, affected, active rates, the number of people affected, dead, recovered). Since these kind of data is dynamically changing at a fast pace, instead of storing into the database, routes bypassed the data to the receiving end of the URL routes. So, the node server we designed acted as a proxy-server for this purpose. Besides we record the daily statistical summary through use of automated scripts.\\
    Another type of route was for the use of data of category 2 and category 3. These data were almost constant or seldom updated. Category 2 includes the prediction based data that were seldom updated and category 3 includes(but not limited to) various information about countries such as area, population, country code, etc.
    Our node server was used to interface with the database. We created a connection pool to connect to the postgresql database in use (both development and deployment). The connection pool system is more efficient in terms of concurrency. The connection pool only connects to the database as needed, such as only during handling a certain database query when a certain URL route or endpoint is called, and then immediately dropping the connection when the query returns with a response(correct or erroneous) in both the cases. After the database response, the data is filtered and formatted in a specific way that would be suitable to the front end part of our web application. Then after data processing, a suitable JSON response was formulated and sent to the receiving side of the URL routes or endpoints.
\subsubsection{Front End} 
    Using React and Material-UI, we developed our user interface. At first, when the user accesses our web-page, we send the new accessed information to back end using the `Post' method of `Axios' library. This information lets us keep track of the number of total visits to our website and using other tools, allows us to locate the country from which the user accessed. The next task is to get the data from back end and display them, for this purpose, we called a JavaScript function `Fetch' with appropriate API route in the respective React components. There are three main elements by which we displayed our data to the users. They are as follows:
    \begin{itemize}
        \item \textbf{Chart:} React-chartjs-3 plugin was mainly used to draw the charts. Among different kind of charts, Line-Charts and Bar-Charts were depicted. The charts were used to represent predictions related to pandemic and analysis of different factors and affect of those factors. Some of the graphs were parameterized according to country using "InputForm" of Material-UI, so the viewers can get a overview of comparison of a particular country with the global condition. Besides, chart annotations were used to specify a particular point in the graphs and colour consistency with the other components were also maintained in the charts.  
        
        \item \textbf{Table:} For the purpose of portrayal of tables, Material-UI-a plugin which is highly compatible with React-js was used. The tables were mainly used to show predictions and latest global condition and conditions of different countries. The data collected from back end were displayed. For the convenience of the users, the headers were made of "Button" component, so that, with the click action, tables were sorted in ascending or descending order according to the values of a particular column.
                
        \item \textbf{Map:} React-leaflet, which is a React wrapper library for the original Leaflet.js library, was used to develop the map. Displaying the live data of affected and dead by countries word-wide, there were two types of map to be displayed: 1) world street map, 2) choropleth map. The tiles for world street map was generated from Esri tile-server. World street map was necessary to visualize the country-wise spreading and choropleth map was necessary for visualizing spreading through others contexts such as, temperature, humidity, population density, etc. Choropleth map was created from a file containing geojson data. Also we needed to include legends to the choropleth map, for which React-leaflet-control was used. Other material-ui components were used to help the navigation of users, such as: Tabs, Radio-buttons, etc. 
    \end{itemize}

\subsection{Data Pre-processing}
Data pre-processing is an important step which involves transforming raw data into an understandable way to interact. We collected our data from a number of different sources. We followed some pre-processing steps that are brefely summarized below: 
    \begin{itemize}
        \item We fetched data from various APIs, and then we needed to process the data as required for back end and analysis. For instance, we got daily affected, dead, recovered data for each country through an API, but for some countries like Australia, there were province-specific data. Thus we added up all the province-specific data for those countries and served it. These data were then used for predictive analysis purposes. Similarly, we needed to process some data so that it could be pushed to the database.
        \item In order to store data in an efficient way and also to redirect some data to the front end part of our web application directly without storing them(live data about various countries), we needed to process a huge amount of data in the server side of our web application. The API we used to collect the data provided us with lots of superfluous and unnecessary information along with the information we required to store in the database of our application for further experimentation and also to bypass the data to the front end so that they can safely use and visualize the data. So proper processing of data was necessary at every step. Processing data includes filtering data, merging certain data with other parts, calculation of various aspects such as rate calculation (death, active and recovered), calculation of user specific time zone, user IP tracing and API calls for getting location specific information, removing some portion of specific data, filling out some portion of missing data from the database, creating suitable JSON responses so that the front end part of our web application can suitably use the formats to display and visualize the data and also creating suitable structures so that we can easily insert to the database the data we want to store. These are some of the few data processing steps we followed to handle the data we obtained from various API sources.
    \end{itemize}

\subsection{Challenges during Data Collection}
One of the major challenges we faced during fetching data from various API is the instability of API. It was necessary to use various API because a single API does not provide all the information we needed to store to the database, visualize on the front end and also do experiments with them. The first major API failure happened on 22/04/2020. We were using a REST API provided by Microsoft for fetching live data related to affected, dead and recovered persons of the pandemic. Besides we were also using the data in various aspects of our web application. The API was made unauthorized due to reasons unknown and thus we could not access the data anymore. We quickly addressed the problem and solved it using another API \cite{covid-19-API} which seemed to have good documentation and useful endpoints that would benefit us. We rewired each endpoint in our own back end so that our routes can be used to fetch reliable data by both the front end part and also for further research and experimentation.\\
Then there was another issue( around 28/04/2020 ) with a certain content of this REST API. At times, the Global Live Data about affected, dead, recovered seemed to show outdated values and thus it seemed to reduce the integrity of data of our web application and this would be troublesome to work with. We switched to a certain new API \cite{mathidro} for these information particularly. This API was consistent with the John Hopkins University website \cite{johnhopkins} and provided real-time Updates on an hourly basis.\\
There was another small issue with another specific data, namely "Country-wise Live Update". For instance there were some missing values for some important countries. This happened at random times. As we were storing daily summaries of each country of the world(affected, dead, recovered, newly affected, newly dead, newly recovered) in our database, we filled up the missing data from the database.\\
Another major challenge we faced is with persistent data storage. We started out using the Mysql database and wrote the database queries and necessary procedures. However during deployment we saw that the initial cloud service\cite{heroku} supports postgresql for non-verified versions of the service. So we resolved the issue quickly by rewriting the database handling portion of our application so that it is compatible with the postgresql database. We needed to reformat the CRUD operations, particularly the search and insertion queries. This resolved the issue of database compatibility, but one large issue still remained. It was that of using a development database. We were using a local database for data storage and operations. In spite of using github for version control, we still had the issue of databases, as we updated the database with various information, we needed to dump the whole database file and use the latest dump to work with the newly inserted information. The issue was resolved as we switched to a cloud version of postgresql(elephant-sql)\cite{elephantsql} not long after deployment. This provided ease of use for both development and deployment as we could both update and use the database at the same time. Although this way out provided some challenges too, for instance, the free tier of cloud database(elephant-sql) \cite{elephantsql} did not allow more than five concurrent connections and also had limited storage (20 MB). As our collected data was still small, we did not face the issue with database storage. But we needed to be cautious to handle five concurrent connections at once. So we used a connection pool to handle database connections in our application. The benefit of using a pool is that the connection is only maintained when we do a query from our application, and as soon as the query returns a response, the connection is cut off. And this in turn provides safety with connection concurrency with the database.\\
We first deployed our web application to Heroku \cite{heroku}, a hosting service with some limiting features but quite easy to use. Heroku had some serious issues with the free tier also known as the hobbiest version. One of the major one was if the web application is left unaccessed for about 20 minutes, then internally the website is put to sleep automatically. This in turn created unbearable long response time if a user accessed the website just after it was put to sleep. Besides, we were using cloud databases for deployment purposes too, which not being part of Heroku, required quite some time to fetch data from database, then visualize and use the data.\\
However, we resolved all these issues by switching to AWS \cite{aws} and using their dedicated RDS (Relational Database Service).\\
Our server is hosting our website from Mumbai, south region. Switching to AWS server was quite challenging as we had to study the basics of cloud computing and configuring our server as needed. Our website is deployed in a virtual machine (Ubuntu 20.04 LTS) called "EC2 instance". At the very first time we were facing problems generating production build. Production build has compressed version of our JavaScript code, so this makes rendering of file on end user's browser very quick and performance enhancing. We realized that the RAM of our "EC2 instance"(Virtual machine that is serving our website) was not enough to run production build. So, we scaled up our instance's RAM size and could deploy our website successfully.
\subsection{Analysis}

Data analysis is defined as a process of cleaning, transforming, and modeling data to discover useful information. Our data analysis process includes Trend analysis and Statistical analysis methods.

\subsubsection{Trend Analysis}

Trend analysis \cite{trend} is the widespread practice of collecting information and attempting to spot a pattern. Our trend analysis process includes calculations of COVID-19 cases rates, plotting of trends and analysing on the plots for trends. Calculations dispense the percentage of death rate, active rate, recovery rate with respect to affected cases as well as death rate and active rate with respect to recovered cases. We have also calculated death rate, active rate, recovery rate with respect to population per million and area per square kilometre. We have first categorized some countries according to large population, rising affected cases, mostly affected cases and observed how trends of these countries draw an impact on global death rate, active rate and recovery rate.\\ 
The equations for analyzing death rates are -
\begin{equation}
\begin{aligned}
Death\ rate(\%)\ with\ respect\ to\ affected & = \frac{Number\ of\ Deaths}{Number\ of\ Affected}\times 100\\
Death\ rate(\%)\ with\ respect\ to\ recovered & = \frac{Number\ of\ Deaths}{Number\ of\ Deaths + Number\ of\ Recovered}\times100\\
Death\ rate\ with\ respect\ to\ population & = \frac{Number\ of\ Deaths}{Number\ of\ Population(per\ million)}\\
Death\ rate\ with\ respect\ to\ area & = \frac{Number\ of\ Deaths}{Area(per\ 100\ sq.Km)} 
\label{eq:d}
\end{aligned} 
\end{equation}

The equations for analyzing active rates are -
\begin{equation}
\begin{aligned}
Active\ rate(\%)\ with\ respect\ to\ affected & = \frac{Number\ of\ Actives}{Number\ of\ Affected}\times 100\\
Active\ rate(\%)\ with\ respect\ to\ recovered & = \frac{Number\ of\ Actives}{Number\ of\ Actives + Number\ of\ Recovered}\times100\\
Active\ rate\ with\ respect\ to\ population & = \frac{Number\ of\ Actives}{Number\ of\ Population(per\ million)}\\
Active\ rate\ with\ respect\ to\ area & = \frac{Number\ of\ Actives}{Area(per\ 100\ sq.Km)}
\label{eq:a}
\end{aligned}
\end{equation}   
The equations for analyzing recovery rates are -
\begin{equation}
\begin{aligned}
Recovery\ rate(\%)\ with\ respect\ to\ affected & = \frac{Number\ of\ Recovered}{Number\ of\ Affected}\times100\\
Recovery\ rate\ with\ respect\ to\ population & = \frac{Number\ of\ Recovered}{Number\ of\ Population(per\ million)}\\
Recovery\ rate\ with\ respect\ to\ area & = \frac{Number\ of\ Recovered}{Area(per\ 100\ sq.Km)}
\label{eq:r}    
\end{aligned}
\end{equation}
    
\subsubsection{Statistical Analysis}

Statistical data analysis \cite{statistic} encompasses collection, analysis, interpretation, presentation, and modeling of dat. Our statistical analysis process includes analysis on COVID-19 datasets to find out correlation of spreading of COVID-19 with other contexts.\\
Before we looked at correlation method, we defined a dataset which we used to test the method. For finding correlation between spreading of COVID-19 and temperature/humidity, we took country-wise daily affected cases as dependent variable Y and maximum temperature/humidity five days prior \cite{article} to that day as independent variable X from the time series dataset. To find whether countries having high pollution index or high food security index affects the number of total affected cases or not, we took total affected cases per country as dependent variable Y and food security/pollution index as independent variable X. To know how healthcare system of a country accelerates total recovered cases of a country, we took total recovered cases as dependent variable Y and country-wise healthcare index as independent variable X. To observe dependency of death rate on total number of population tests, we referred death rate (Total number of deaths/Total number of affected) as dependent variable Y and total number of population tests as independent variable X. Scatter plot of the two variables has been created to see whether there is any existing trend or not.\\
Before we took a look at calculating some correlation scores, we first looked at an important statistical building block, called covariance \cite{covar}.The calculation of the sample covariance is as follows:
\begin{equation}
    cov(X, Y) = \frac{(sum(x-mean(X)) \times (y-mean(Y)))}{(n-1)}
\end{equation} \\  
here, the use of the mean in the calculation suggests the need for each data sample to have a Gaussian or Gaussian-like distribution. A problem with covariance is, as a statistical tool alone it is challenging to interpret the results. This leads us to the correlation coefficient next.\\
We have determined correlation coefficient \cite{correl} as a statistical measure of the strength of the relationship between the relative movements of two variables. The values range between -1.0 and 1.0 with 0 implying no correlation. Positive correlations imply that as x increases, so does y. Negative correlations imply that as x increases, y decreases.  We have used Pearson's correlation analysis \cite{wikip}, Spearsman's rank correlation analysis \cite{wikis} and Kendall's rank correlation analysis \cite{wikikend} to find out if there is significance of temperature and humidity in the spread of COVID-19. Also we have run the tests to find out if there is any impact of food security index in encouraging the COVID-19 cases, if pollution exacerbates the virus, if number of recovered cases depend on heath index of a country and lastly how death rate changes with the number of population tests.
The calculation of the sample Pearson correlation coefficient is as follows:\\
The coefficient is often denoted by the lowercase r.\\
\begin{equation}
    Pearson's\ rank\ correlation\ (r) = \frac{covariance(X, Y)}{(stdv(X) \times stdv(Y))}
    \label{eq:pearson}
\end{equation}\\
here, $stdv$ stands for standard deviation. The use of mean and standard deviation in the calculation suggests the need for the two data samples to have a Gaussian or Gaussian-like distribution. So, this procedure cannot be used for data that does not have a Gaussian distribution. As a result, we have used rank correlation \cite{wikirank} methods as an assurance of better analysis. The calculations for two renowned rank correlation methods are:\\    
\begin{equation}
    Spearman's\ rank\ correlation\ (\rho) = 1 - \frac{6\sum d_i^{2}}{n\times(n^{2}-1)}
    \label{eq:spearman}
\end{equation}\\
here, \textbf{d\textsubscript{i}} is modulus of  different between rank of X\textsubscript{i} and Y\textsubscript{i}, \textbf{n} denotes the total number of samples.\\
\begin{equation}
   Kendall's\ rank\ correlation\ (\tau) = \frac{C-D}{\frac{n\times(n-1)}{2}}
   \label{eq:kendall}
\end{equation}\\
here, \textbf{C} and \textbf{D} denote number of concordant pairs and number of discordant pairs respectively, \textbf{n} denotes total number of samples.\\
We calculated Pearson's correlation in Python using the pearsonr() SciPy function \cite{pythonp}, spearman’s rank correlation in Python using the spearmanr() SciPy function \cite{pythons} and Kendall’s rank correlation coefficient in Python using the kendalltau() SciPy function \cite{pythonk}.\\

\subsection{Analytics}
    Data analytics is the science of scrutinizing raw data in order to make conclusions about that information. Our data analytics process incorporates analyzing COVID-19 datasets and predicting future of COVID-19 with the numbers of affected, dead, recovered per country. 
    In order to perform the prediction, we started off with a compartmental model of epidemiology, the basic \textbf{SIR} model \cite{wikiSir}. This model breaks down the population into three compartments: \textbf{S} for the number of Susceptible, \textbf{I} for the number of Infectious, and \textbf{R} for the number of Removed Individuals. A set of three ordinary differential equations mentioned below is the formulation of the model.

    \begin{equation}
    \begin{aligned}
    \frac{dS}{dt} & =  - \beta S(t) I(t)  \\
    \frac{dI}{dt} & =  \beta S(t) I(t) - k I(t)  \\
    \frac{dR}{dt} & =  k I(t) . 
    \end{aligned}        
    \end{equation}

    
    where \textbeta~is the transmission rate of the infection and \textlambda~is the removal rate of the infected individual.\newline
    This model assumes constant population throughout the time of the epidemic. The model also assumes homogeneous mixing of people, no reinfection of recovered individuals, age, sex, social status, and race independence of being infected. To address these issues, at first we broke down the removed compartment into two compartments: \textbf{R}ecovered and \textbf{D}ead. Then we had the basic \textbf{SIRD} model which can address the difference of the recovered and dead individuals. To address the regular number of births and deaths irrespective of the epidemic, we introduced birth rate(\textmu) and death rate(\textGamma) to the model to make it compatible with vital dynamics. To address the issue of reinfection, we used a time dependent reinfection rate(\textzeta(t)) \cite{Victor:2020}. Since the recovered individuals can be reinfected, we added the relevant portion of them to the susceptible compartment. The relevant portion was determined by the reinfection rate. The formulation of our revised model was      
    
    \begin{eqnarray}
    \begin{aligned}
    \frac{dS}{dt} & =  - \beta S(t) I(t) + \mu N - \Gamma S(t)  \label{eq2a} + \zeta (t) \Gamma R(t)\\
    \frac{dI}{dt} & =  \beta S(t) I(t) - (\lambda_d + \lambda_r) I(t)  - \Gamma I(t) \label{eq2b}\\
    \frac{dR}{dt} & =  \lambda_r I(t)  - \Gamma R(t) \label{eq2c}\\
    \frac{dD}{dt} & =  \lambda_d I(t) - \Gamma D(t) . \label{eq2d} 
    \end{aligned}
    \end{eqnarray}
    
    where N is the total population. \newline
    Finally, in order to interpret the lockdown and social distancing measures, the population was aggregated into M = 16 groups. The aggregation was done with respect to a 5-year age interval, ranging from individuals aged from 1 year old to individuals aged 80 years old. Each group represents a class comprising of corresponding aged individuals.
    Representing a class as $i$, we get the total population size \cite{Towers:2012} 
    \begin{equation}
         N = \sum_{i=1}^{M}{S_i(t) + I_i(t) + R_i(t) + D_i(t)} = \sum_{i=1}^{M}{N_i}
    \end{equation}
    
    where $S_i$, $I_i$, $R_i$, and $D_i$ is number of susceptible, infectious, recovered, and dead individuals of the class $i$.\newline
    After dividing the population into M classes, we used the social contact structures between all possible classes. We considered these mixing patterns of all possible classes at home, school, working place, and other locations  \cite{10.1371/journal.pcbi.1005697}. Thus, we were able to determine the number of people of a certain class coming into the contact of another class at different places. \newline
    Let \textbf{C} be the contact matrix whose each element ${C_{ij}}$ represents the average number of contacts made per day by an individual in class $i$ with an individual in class $j$. 
    We can partition $C_{ij}$ into $C$\SPSB{$H$}{$ij$}, $C$\SPSB{$S$}{$ij$}, $C$\SPSB{$W$}{$ij$}, and $C$\SPSB{$O$}{$ij$} which represent the contact structures at home, school, working place, and other locations respectively. Thus, we get \cite{10.1371/journal.pcbi.1005697} 
    \begin{equation}
        C_{ij} = \delta_1C\SPSB{H}{ij} + \delta_2 C\SPSB{S}{ij} + \delta_3 C\SPSB{W}{ij} + \delta_4 C\SPSB{O}{ij}
    \end{equation}
    
    where \textdelta$_1$, \textdelta$_2$, \textdelta$_4$, and  \textdelta$_4$ are the indicators of mixing patterns at different places. These constants were varied according to different lockdown and social distancing states \cite{Singh:2020}.

     Now, assuming all the infections are symptomatic, the formulation of our revised model is \cite{Towers:2012} 
    
    \begin{eqnarray}
    \begin{aligned}
    \frac{dS_i}{dt} & =  - \beta\sum_{j=1}^{M}{C_{ij}\frac{I_j}{N_j}} S_i(t) I_i(t) + \mu N_i - \Gamma S_i(t)  \label{eq2a} + \zeta (t) \Gamma R_i(t)\\
    \frac{dI_i}{dt} & =  \beta\sum_{j=1}^{M}{C_{ij}\frac{I_j}{N_j}} S_i(t) I_i(t) - (\lambda_d + \lambda_r) I_i(t)  - \Gamma I_i(t) \label{eq2b}\\
    \frac{dR_i}{dt} & =  \lambda_r I_i(t)  - \Gamma R_i(t) \label{eq2c}\\
    \frac{dD_i}{dt} & =  \lambda_d I_i(t) - \Gamma D_i(t) . \label{eq2d}
    \end{aligned}
    \end{eqnarray}

\section{RESULTS}
Analysis with the processed data led us to some interesting inferences. Our analysis encompassed both trend analysis for some specific countries and also some statistical tests to discover if any kind of correlation is prevalent among the data and the factors. Besides, some enticing observations regarding the predictions based on the revised \textbf{SIRD} model was obtained. The results are illustrated below.

\subsection{Death Rate, Active Rate, Recovery Rate Analysis}
On the basis of data collected from API \cite{covid-19-API} \cite{worldbank} \cite{population} from Jan 21 to May 13 and calculations done by using equations \ref{eq:d}, \ref{eq:a}, \ref{eq:r}, we can draw some conclusions which are:\\
\subsubsection{Death Rate with Respect to Affected}
     
    \begin{figure}[!h]
    \centering
    \def\svgwidth{\columnwidth}
    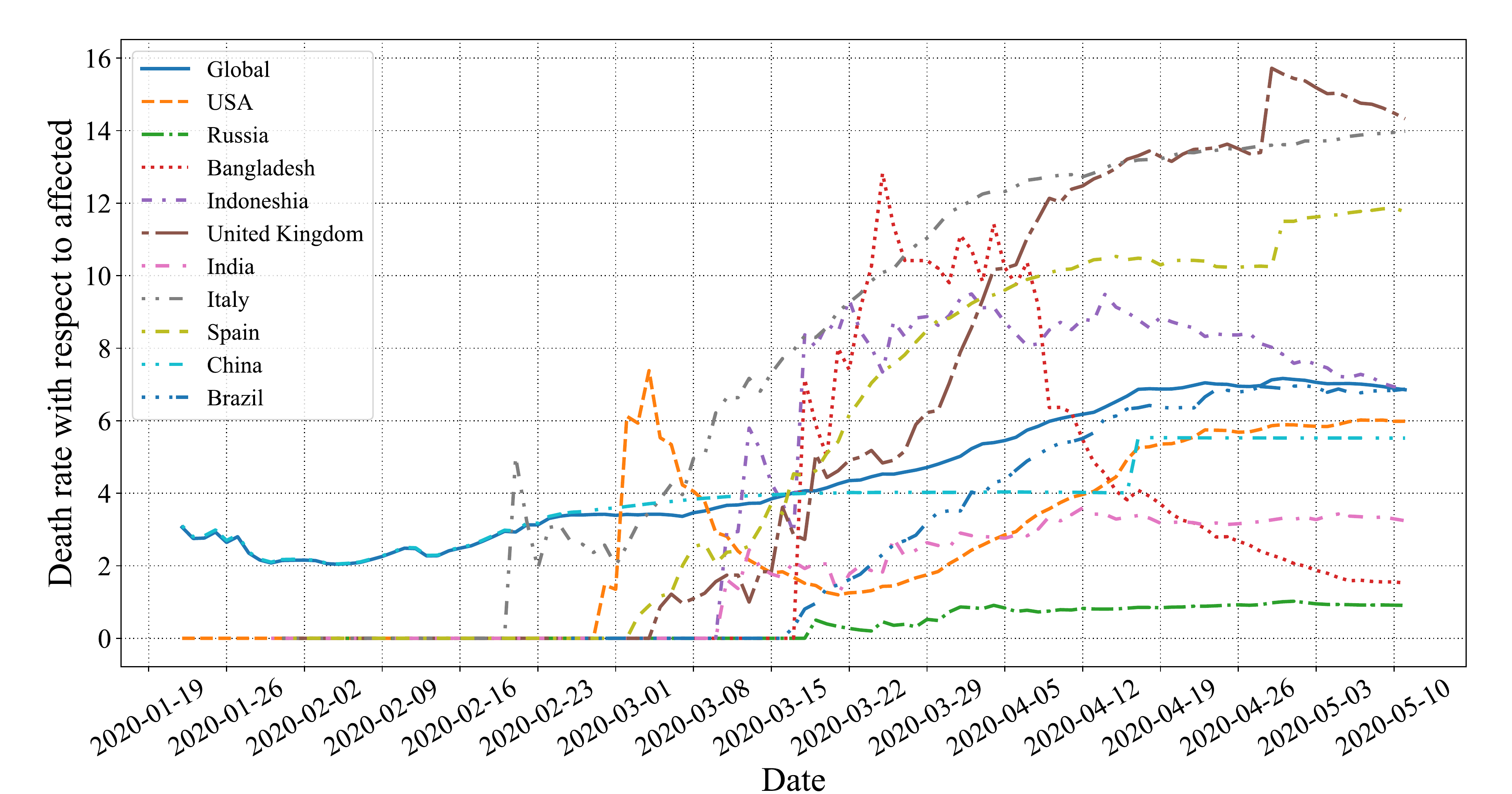
    \caption{\textbf{Death Rate with Respect to Affected}}
    \label{fig:da}
    \end{figure}
    According to Figure: \ref{fig:da}, we have drawn some observations here:\\
    The Global death rate curve is on a constant rise with an average slope of around 0.09 which varies according to date range. The slope went almost flat since mid-April with a negligible slope. Some notable countries with a high death rate compared to Global death rate are Spain, Italy, Indonesia, United Kingdom. Some notable countries with a medium death rate compared to the Global death rate are  China, Bangladesh, Brazil, and the United States of America. The curve of the United States of America and Bangladesh was high around March and April (March 02 - March 06 for the United States of America and March 18 - March 22 for Bangladesh). Some notable countries with a low death rate compared to the Global death rate are Russia, India. Some fluctuations were observed for the following countries which are the United States of America, Indonesia, and Bangladesh. Around March, the death rate of these countries increased drastically, like for Bangladesh it reached around 13\%.
    
\subsubsection{Death Rate with Respect to Recovered}
    
    \begin{figure}[!h]
        \centering
        \def\svgwidth{\columnwidth}
        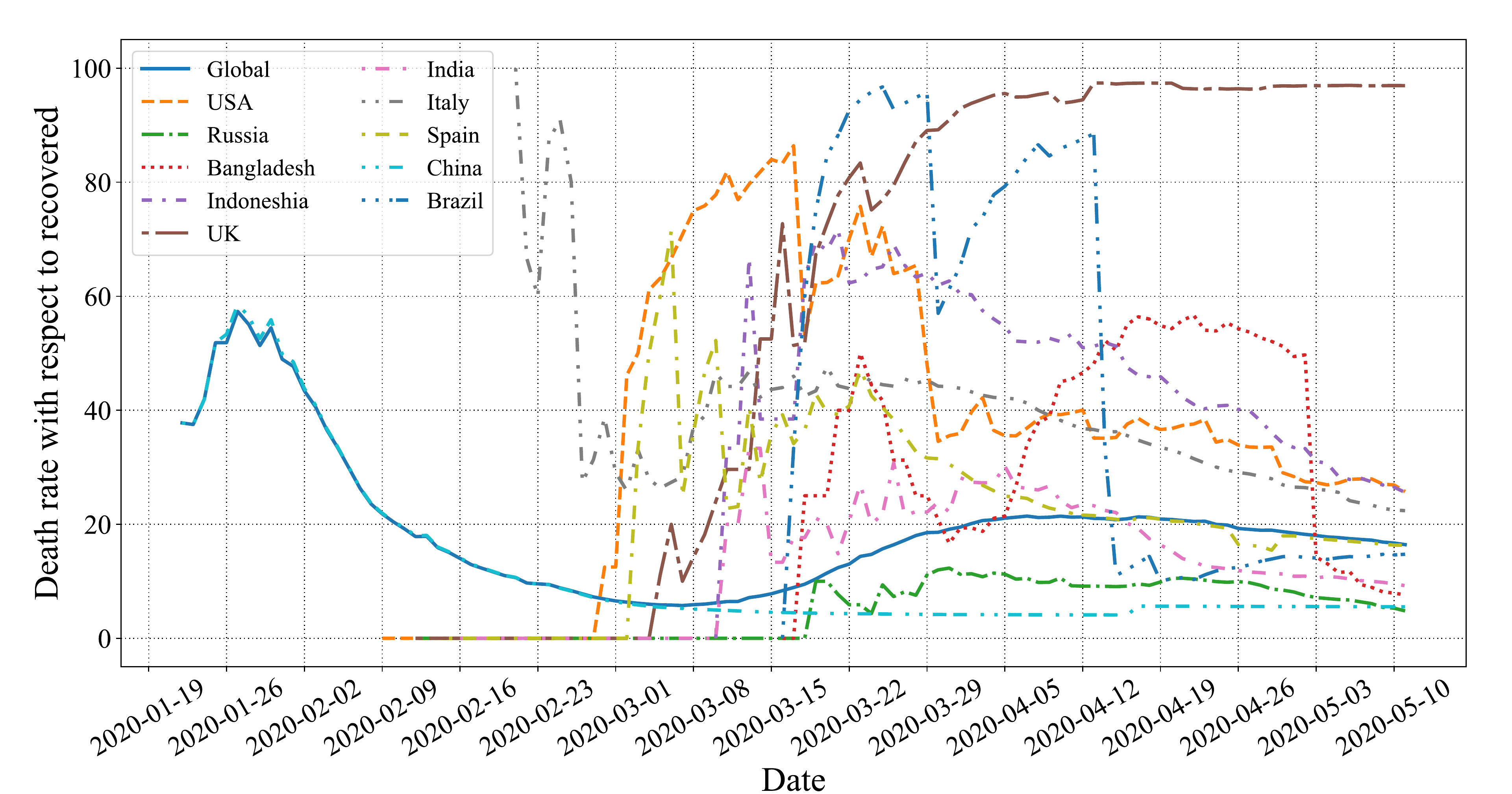
        \caption{\textbf{Death Rate with Respect to Recovered}}
        \label{fig:dr}
    \end{figure}
    According to Figure: \ref{fig:dr}, our key findings are:\\
    Initially, there was a rise in the Global death rate with a slope of 2.3 from Jan 21 to Jan 27, and following that, there was a decrease in death rate with a slope of 2 from Jan 28 to February 02. After that, the slope somewhat became constant with an average slope of 0.2. Some notable countries with a high death rate compared to Global death rate are the United States of America (especially in March, highest = 100\%), Italy (highest = 100\%), Indonesia (highest = 72\%), Bangladesh (highest = 56\%), and the United Kingdom (highest = 97\%). Some notable countries with a medium death rate compared to the Global death rate are Spain, India, Brazil. For Spain and Brazil, the curve was drastically high around March and April and reached around 97\% and 71\% for Brazil and Spain respectively. China, Russia have a low death rate compared to Global rate.
        
\subsubsection{Death Rate with Respect to Population}

    \begin{figure}[!h]
        \centering
        \def\svgwidth{\columnwidth}
        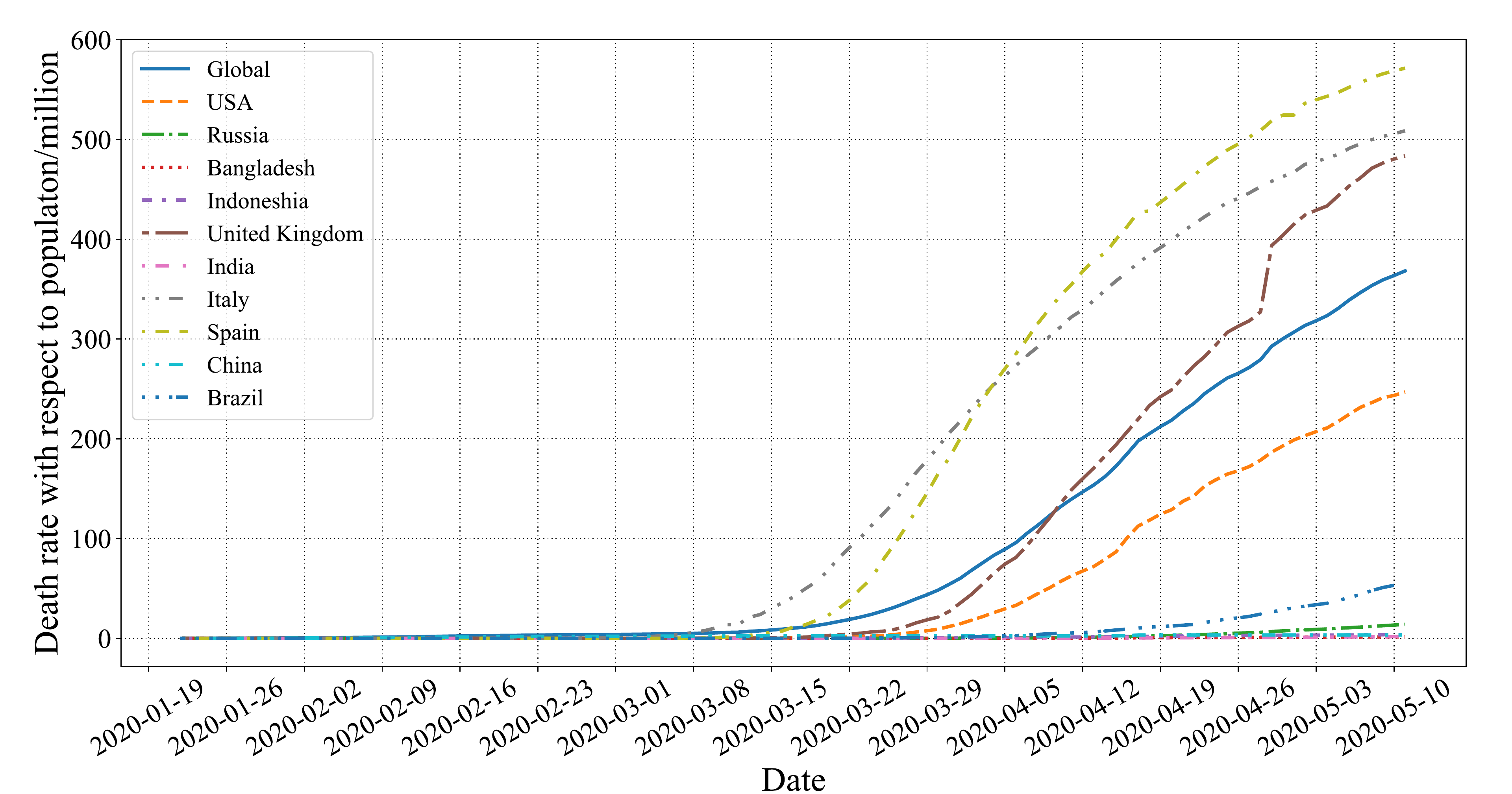
        \caption{\textbf{Death Rate with Respect to Population}}
        \label{fig:dp}
    \end{figure}
    Our observations from Figure: \ref{fig:dp} are given below:\\
    The Global death rate started rising from 14/03/2020 with an average constant slope of 0.3. Some notable countries with a high death rate compared to the Global death rate are the United States of America, Italy, Spain, and the United Kingdom. All of these countries had low death rates but from mid-March, the slope started rising. Some notable countries with a medium death rate compared to Global death includes Brazil. Some notable countries with a low death rate compared to the Global death rate are Indonesia, China, Russia, India, Bangladesh.
        
\subsubsection{Death Rate with Respect to Area}

    \begin{figure}[!h]
        \centering
        \def\svgwidth{\columnwidth}
        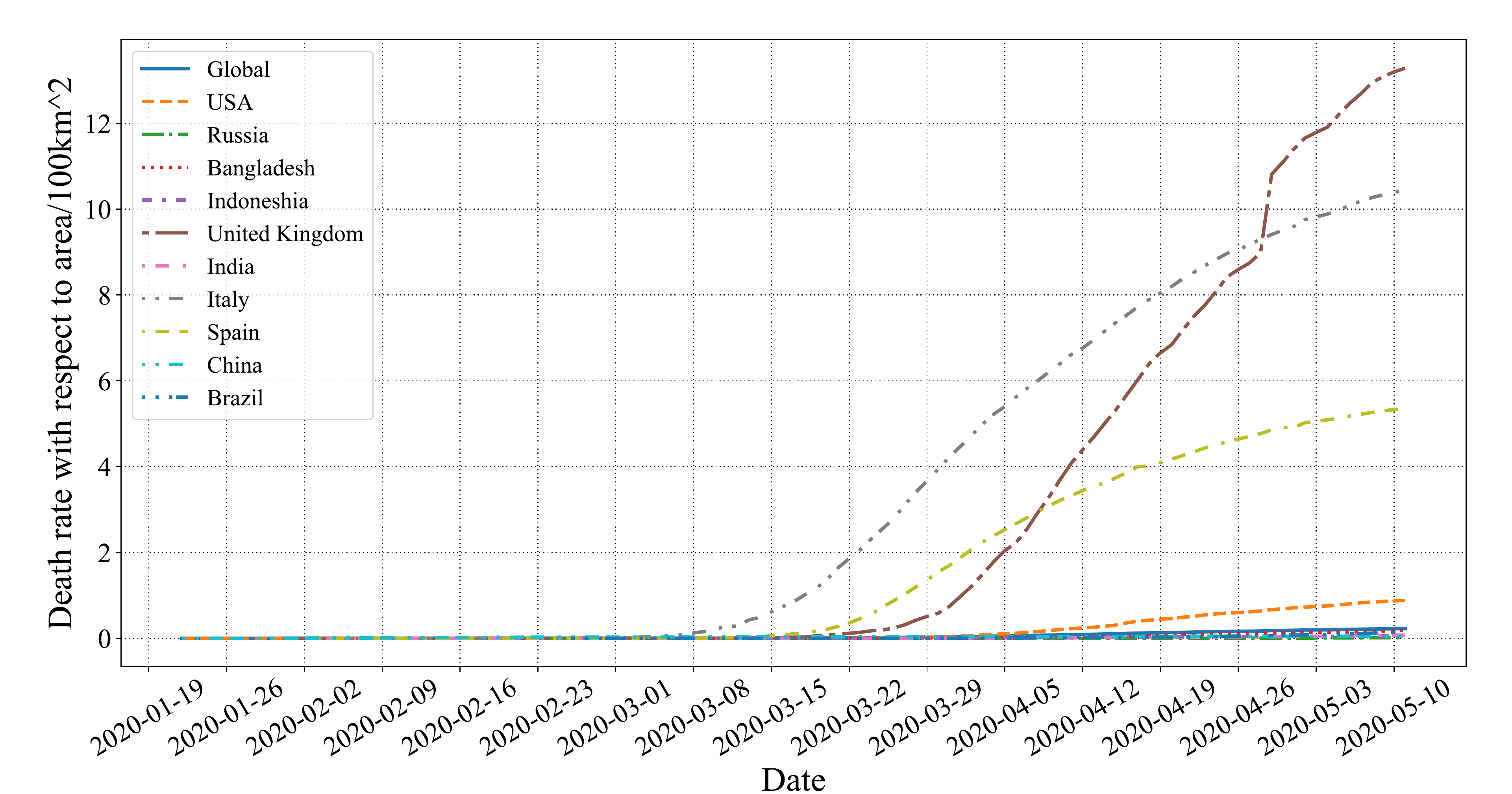
        \caption{\textbf{Death Rate with Respect to Area}}
        \label{fig:dar}
    \end{figure}
    We have noted some findings below from Figure: \ref{fig:dar}:\\
    The Global death rate was nearly constant within this time period and the value never crossed 0.2 with a negligible slope. Some notable countries with a high death rate compared to the Global death rate are the United States of America, Italy, Spain, and the United Kingdom. Countries with a medium death rate compared to the Global death rate are China, Brazil, Bangladesh. Some notable countries having low death rate compared to the Global death rate are Indonesia, Russia, India.

\subsubsection{Active Rate with Respect to Affected}

    \begin{figure}[!h]
        \centering
        \def\svgwidth{\columnwidth}
        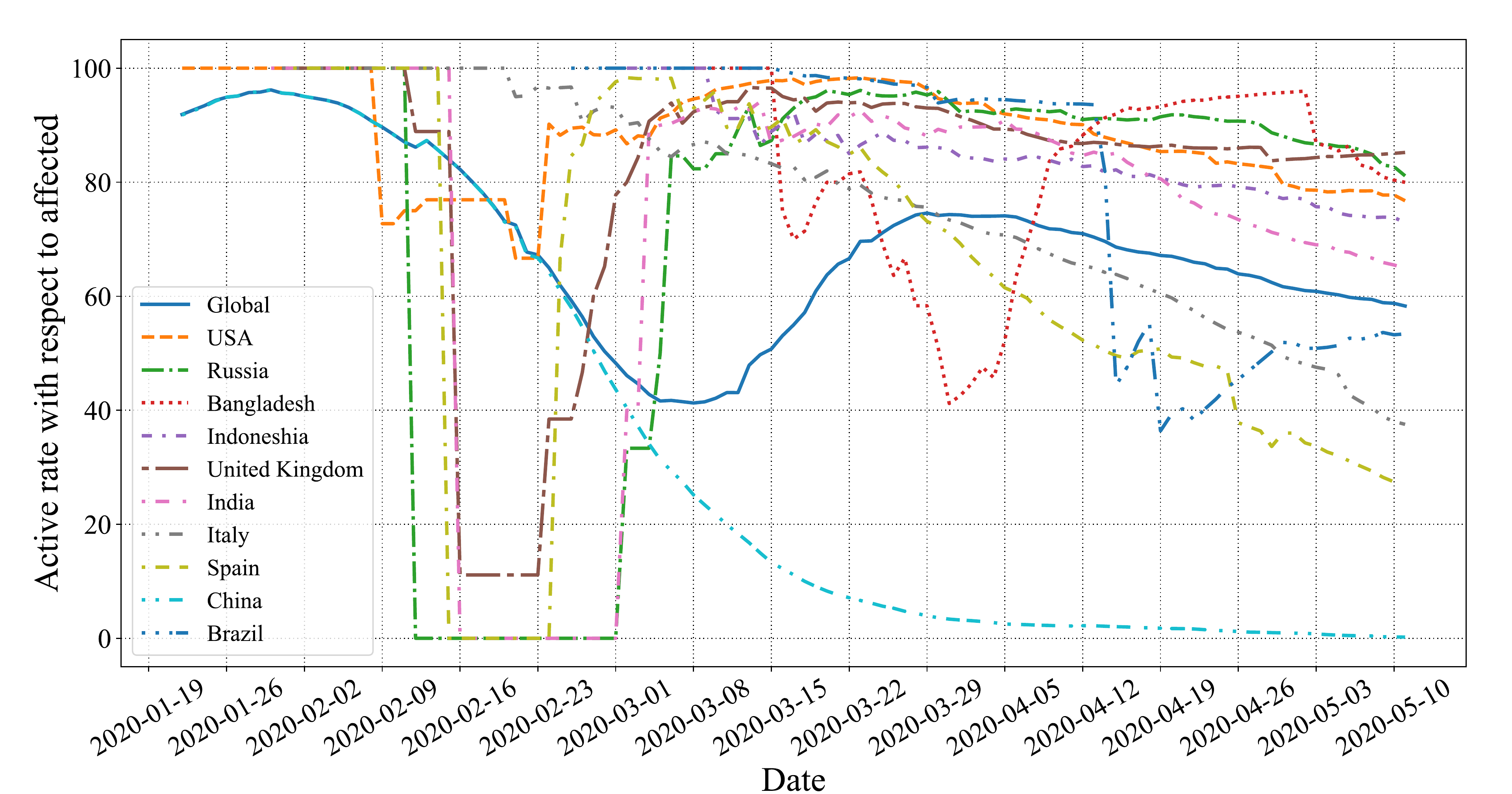
        \caption{\textbf{Active Rate with Respect to Affected}}
        \label{fig:aa}
    \end{figure}  
    From Figure: \ref{fig:aa}, we have summed-up our findings:\\
    Global Active Rate shows a slight upward trend, then a gradual downward trend reaching a minimum value of 41\%. The maximum active rate so far was 96\%, recorded at Jan 29. Some notable countries with a high active rate compared to Global active rate are Russia (highest = 100\%), USA (highest = 100\%), Bangladesh (highest = 100\%), Indonesia (highest = 100\%), United Kingdom (highest = 100\%), India (highest = 100\%). Some notable countries with medium active rates are Spain and Italy. Both the countries show a general downward trend of active rate. China is one of the notable countries to have a lower active rate than the Global active rate, owing to the fact that China has been fighting the pandemic for the longest period of time. Up Until 14/04/2020, the curve of Brazil was well above the Global active rate, showing a higher active rate. Then there was a sudden drop of the active rate from April 04 to April 19. After that, the active rate suddenly increased from April 20 and currently almost converged with the global active rate.

\subsubsection{Active Rate with Respect to Recovered}

    \begin{figure}[!h]
        \centering
        \def\svgwidth{\columnwidth}
        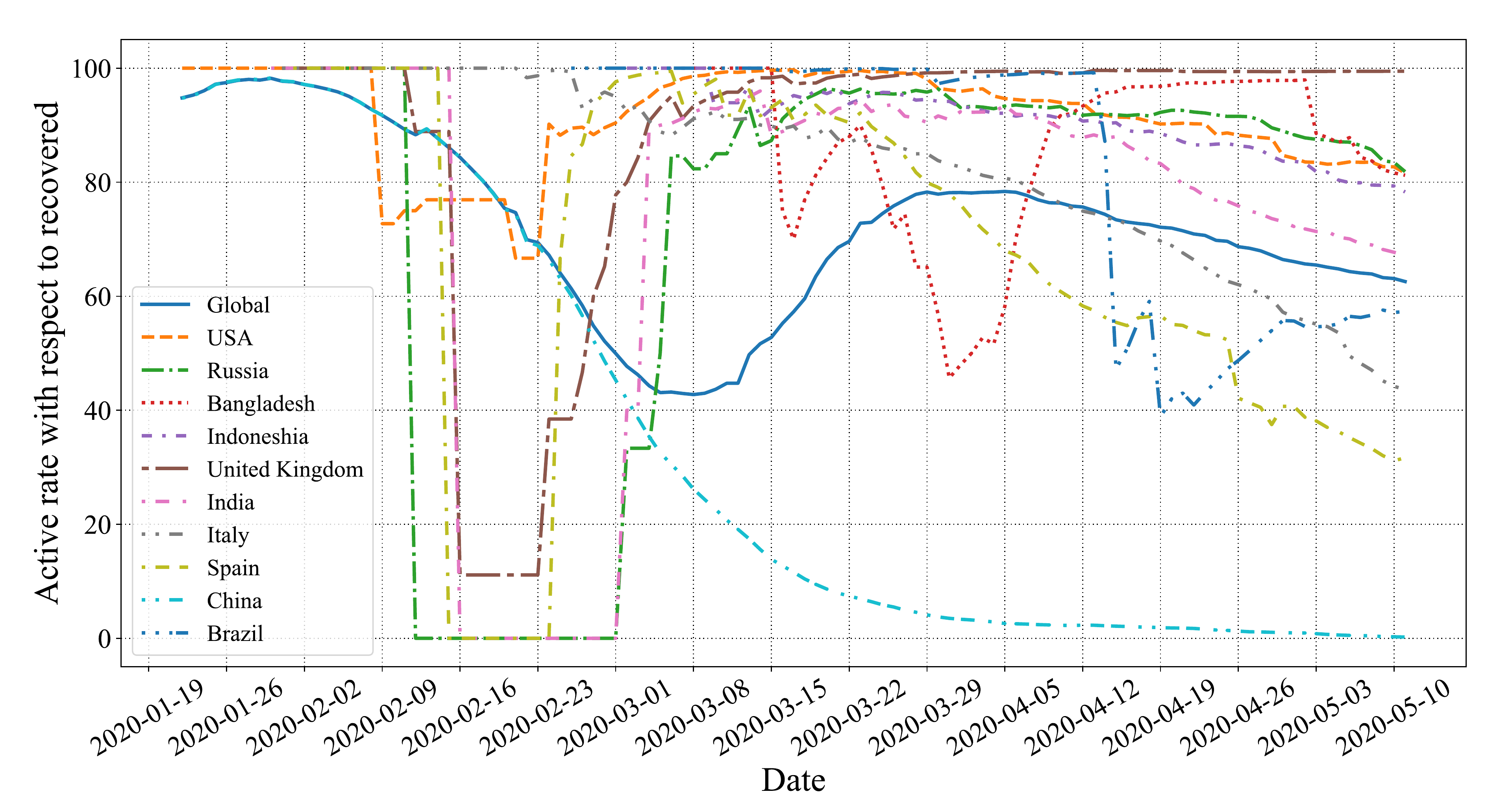
        \caption{\textbf{Active Rate with Respect to Recovered}}
        \label{fig:ar}
    \end{figure}
    We have mentioned our observations from Figure: \ref{fig:ar} below:\\
     Global Active Rate shows a slight upward trend, then a gradual downward trend reaching a minimum value of 42\%. The maximum active rate so far was 98\%, recorded at 29/01/2020. Some notable countries with a high active rate compared to the Global active rate are Russia (highest = 100\%), United States of America (highest = 100\%), Bangladesh (highest = 100\%), Indonesia (highest = 100\%), United Kingdom (highest = 100\%), India (highest = 100\%). One interesting trend for the United Kingdom is that the active rate for this criteria seems to be increasing. For Bangladesh, initially the curve was above the Global active rate, then afterwards the curve was below the Global active rate from March 26 to April 07. Afterwards, it shows a higher active rate than the global one. China possesses a lower active rate than the global active rate, owing to the fact that China has been fighting the pandemic for the longest period of time.
        
\subsubsection{Active Rate with Respect to Population}

    \begin{figure}[!h]
        \centering
        \def\svgwidth{\columnwidth}
        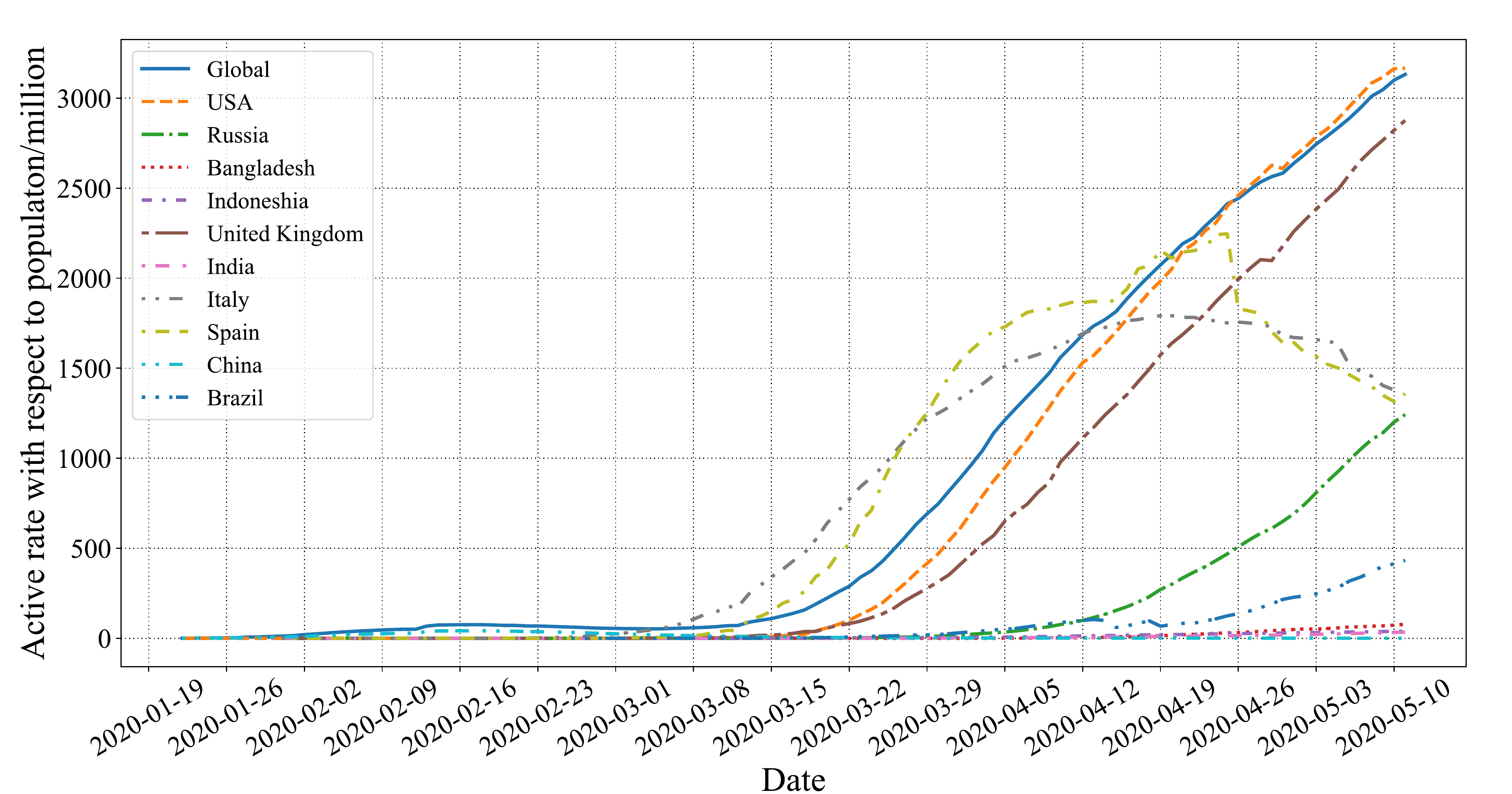
        \caption{\textbf{Active Rate with Respect to Population}}
        \label{fig:ap}
    \end{figure}
    From Figure: \ref{fig:ap}, we have mentioned our key findings:\\
    The Global active rate was below 10 till March 15, then it shows a constant upward trend with a positive slope. The active rate is increasing as the day progresses. Some notable countries with high active rates with respect to Global active rate are United States of America (highest = 3323), United Kingdom (highest = 3005), Spain (highest = 2247), Italy (highest = 1791.7). For the United States of America and the United Kingdom, the curve seems to follow a similar trend in active rate. We again see a similar trend for Spain and Italy. Both the curves show an upward trend with positive slope and then a gradual downward trend. Russia can be considered to have a medium active rate with respect to Global active rate because it seems to be below the Global active rate till 16/04/2020. Then it crossed the Global active rate and shows an upward trend with a positive slope till April 13. For Bangladesh, India, Indonesia, the active rate is well below the Global active rate with a rising trend for the individual countries. For China, it seems to be flattening out after April 22.
    
\subsubsection{Active Rate with Respect to Area}

    \begin{figure}[!h]
        \centering
        \def\svgwidth{\columnwidth}
        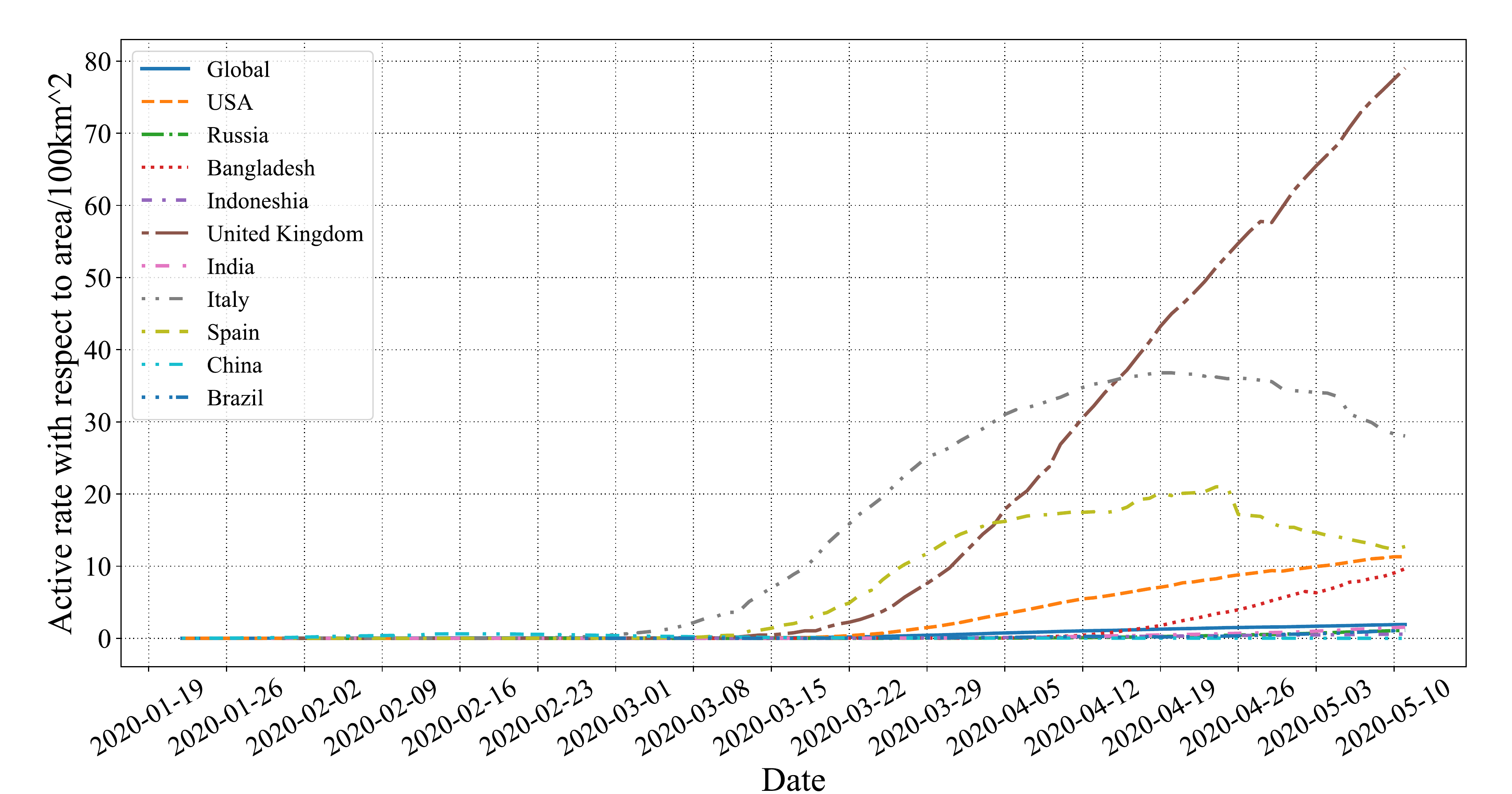
        \caption{\textbf{Active Rate with Respect to Area}}
        \label{fig:aar}
    \end{figure}
    Our major findings from Figure: \ref{fig:aar} are described below:\\
    The Global active rate was somewhat linear till March 15, then it shows a constant upward trend with a positive slope. The active rate is increasing as the day progresses. Some notable countries with high active rate with respect to Global active rate are United States of America (highest = 12), United Kingdom (highest = 83), Spain(highest = 21), Italy (highest = 37). For the United States of America, the value was almost linear till March 18 and for the United Kingdom it was till March 15. Then for both the curves, we see a constant rise of active rate with positive slope. For Bangladesh and China, we observe an interesting trend in the active rate. For Bangladesh, we observe that the curve was below the global curve till April 15. Then Bangladesh crossed the Global active rate and has a much higher active rate with a constant rising trend with a positive slope of the curve. And for China, we see an opposite trend, we see a very high active rate compared to Global active rate from Jan 28 to March 15 with a maximum value of 0.61895 (February 17). Then the active rate was below the global one, and seems to be flattening out after March 24.
    
\subsubsection{Recovery Rate with Respect to Affected}

    \begin{figure}[!h]
        \centering
        \def\svgwidth{\columnwidth}
        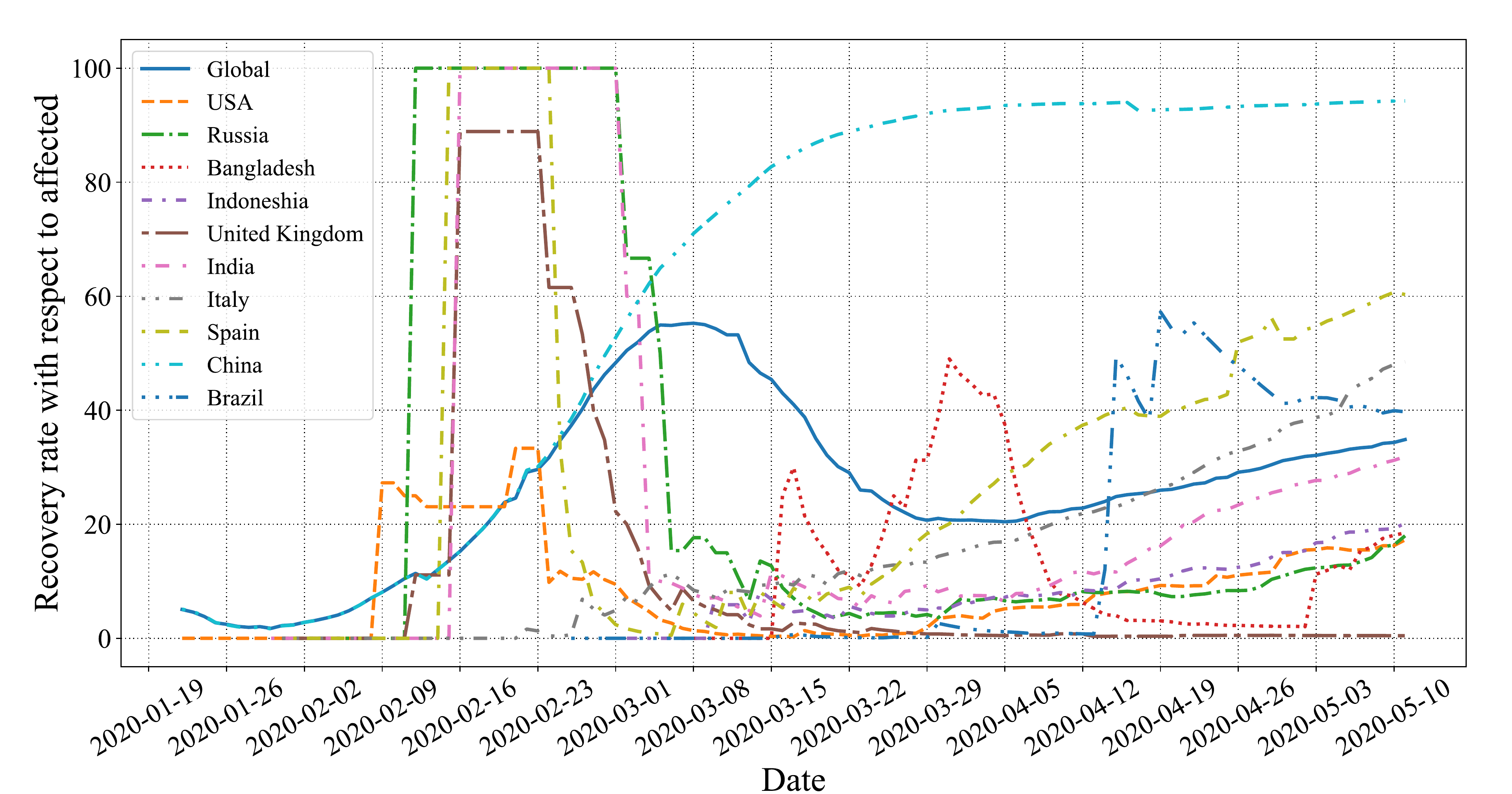
        \caption{\textbf{Recovery Rate with Respect to Affected}}
        \label{fig:ra}
    \end{figure}
    Findings from Figure: \ref{fig:ra} are:\\
    Global recovery rate had a upward value from 2\% to 61\% till March 12. It depicted a downward value from 61\% to 20\% till April 06. Now the value is almost constant. China is showing very high recovery rate with a value ranging from 56\% to 94\% in past 70 days. Italy had a lower recovery rate and now it is rising. Recovery rate of Brazil was very slow till mid-April. Now it is rising significantly. United States of America has been showing lower recovery rate with respect to affected compared to Global recovery rate.

\subsubsection{Recovery Rate with Respect to Population}
     
    \begin{figure}[!h]
        \centering
        \def\svgwidth{\columnwidth}
        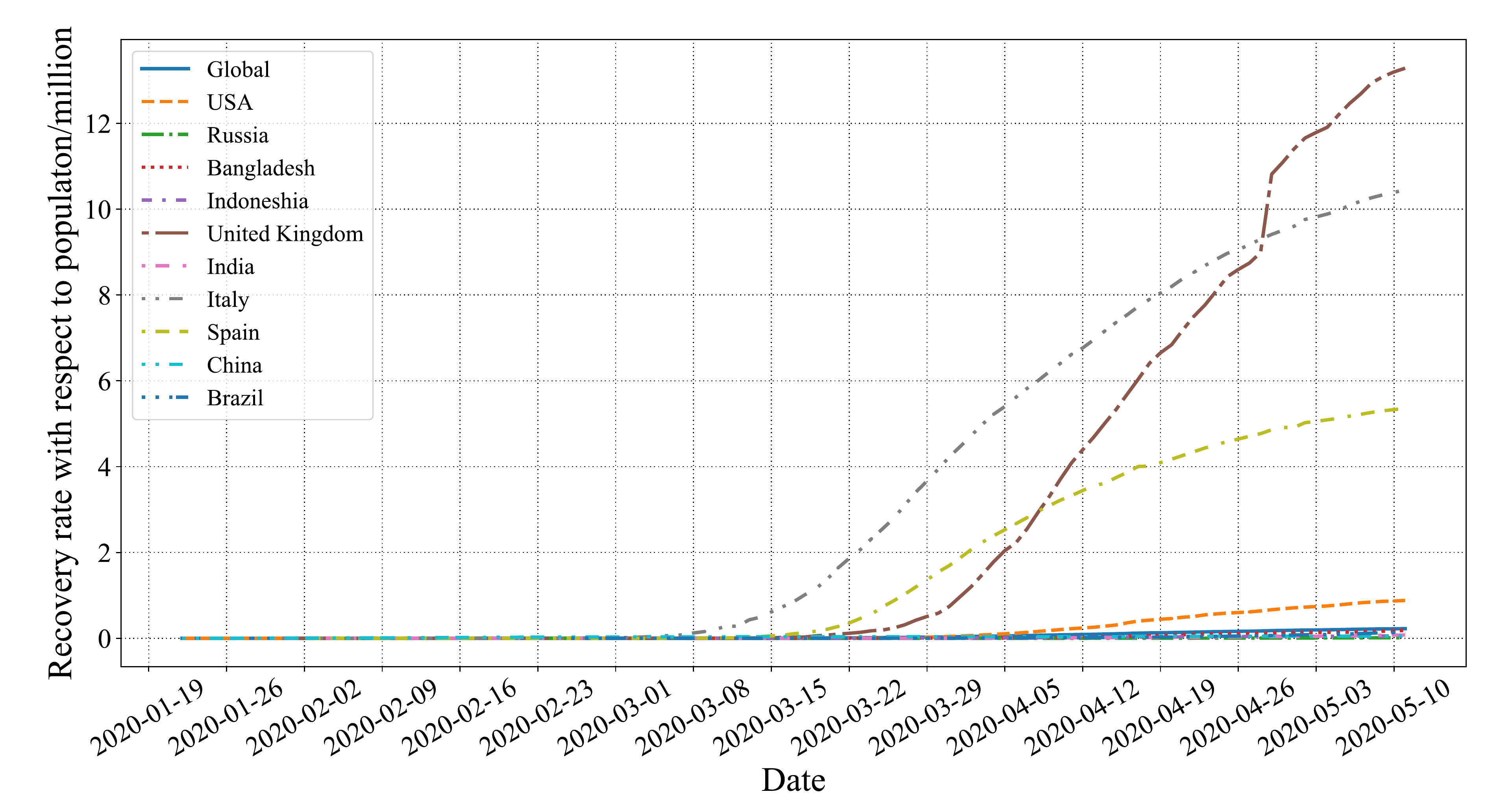
        \caption{\textbf{Recovery Rate with Respect to Population}}
        \label{fig:rp}
    \end{figure}
    Observations from Figure: \ref{fig:rp} are:\\
    Global recovery rate is on a constant rise within this time period and the current value is around 196/M. Italy and Spain have been showing a upward trend within this time period and the current value is around 1900/M and 3080/M respectively. Bangladesh and India have a very low rate of recovery and current value is showing 20/M and 15/M respectively.
    
\subsubsection{Recovery Rate With Respect to Area}
     
    \begin{figure}[!h]
        \centering
        \def\svgwidth{\columnwidth}
        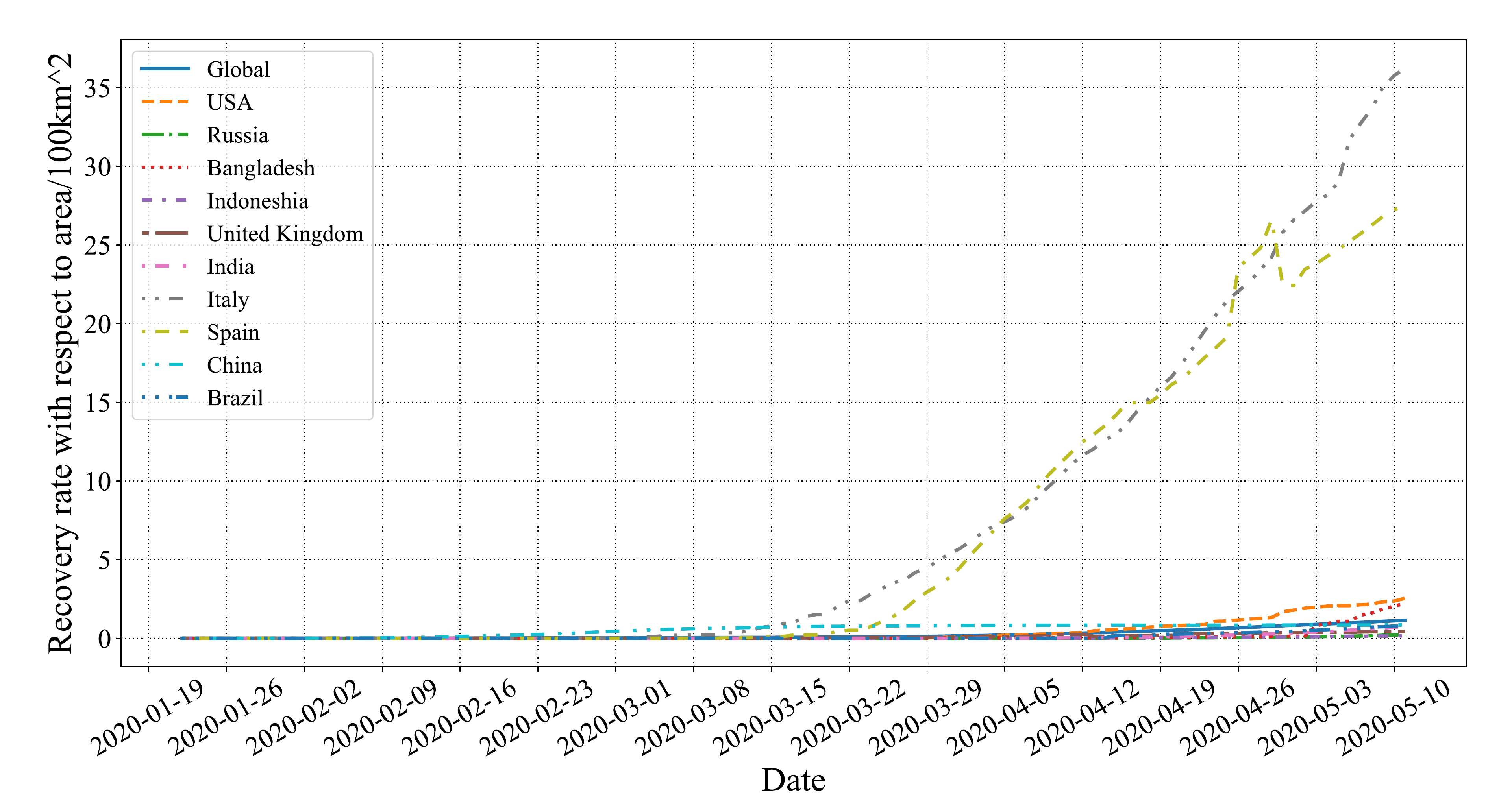
        \caption{\textbf{Recovery Rate with Respect to Area}}
        \label{fig:rar}
    \end{figure}
    We have drawn some observations below from Figure: \ref{fig:rar}:\\
    Global recovery rate is on a constant rise within this time period and the current value is around 1.13/100 sq.Km. Spain and Italy are notable for having higher recovery rate compared to Global rate. China is showing opposite trend of Global recovery rate.\\
    
    The findings help to address the current situation of the pandemic and lead to make public and government decisions on disease surveillance process. High risk areas and populations can be targeted for the intervention of the pandemic. Also, these analyses provide a valuable archive for disease activities for future references and responding to disease outbreaks
     
\subsection{Correlation with Other Contexts}

As we mentioned in section \textbf{2.5}, we have used \nameref{eq:pearson}, \nameref{eq:spearman} and \nameref{eq:kendall} method to find out any correlation among spreading of COVID-19 and environmental factors such as temperature, humidity, pollution index, among recovered cases of COVID-19 and quality of life such as healthcare index, food quality index, among death rate and population tests. To determine whether the correlation between variables is significant, we have compared the p-value to our defined significance level. We used 0.05 as our significance level (denoted as $\alpha$ or alpha). An $\alpha$ of 0.05 indicates that the risk of making a conclusion about existence of a correlation when, actually, no correlation exists is 5\%.\\
P-value $\leq\ \alpha$ : The correlation is statistically significant.
If the p-value is less than or equal to the significance level, then we can conclude that the correlation is different from 0. \newline
P-value > $\alpha$: The correlation is not statistically significant
So, if the p-value is greater than the significance level, then we cannot conclude that the correlation is different from 0.  We present the affected-environmental, recovered-quality of life, death rate-population tests characteristics obtained from the analysis below in detail.
To observe correlation with humidity and temperatures, we have used time series data. Different country has different number of data points as we ignored 0 newly affected cases. We have also collected historical environmental data from Jan 22 to April 01. According to our result of correlations from Table: \ref{tab:table}, we have found that, correlation coefficient for temperature and affected ranges between -0.002 and 0.7 for 114 countries. Since the coefficients showed skewed distribution, we have used median as a parameter of overall correlation coefficient of the whole world. median value of temperature-affected correlation coefficient is 0.09. The value is small and positive which means though daily affected cases of some countries show significance correlation with temperature, the overall impact of it on COVID-19 is negligible. Correlation coefficient for humidity and affected ranges between -0.009 and 0.55 for 114 countries. The median value of humidity-affected correlation coefficient of all the countries is -0.5. The value of is significant and negative which means daily affected cases of some countries show significance correlation with humidity and if humidity increases, a decreasing trend of affected cases is found.The overall impact of it on COVID-19 is neither negligible nor strong enough to draw a conclusion.
To observe correlation with pollution, healthcare, food security and population tests, we used average data of socio-economic factors and latest data of COVID-19 cases. Our latest data for COVID-19 cases for all socio-economic analysis is of May 29th. Every country has one data point for socio-economic analysis. As a result, we have determined overall (median) value of correlation for worldwide result. Healthcare, pollution index data was collected from Numbeo \cite{numbeo} and was last updated on 5 May, 2020. We collected data of 108 countries and 80 countries for pollution index and healthcare index respectively. Data of 113 countries of food security index has been collected from Global Food Security Index \cite{food} and was last updated on November, 2019. Population tests data has been collected from an API \cite{death} and death rate of every single country has been calculated using formula:\begin{equation}
    Death\ rate\ =\frac{\ Total\ number\ of\ deaths}{Total\ number\ of\ affected}
\end{equation}
According to our result of correlations from Table: \ref{tab:table}, we have found that the overall correlation coefficient for pollution and total affected is -0.12. The value is negative and negligible. The overall value of food security-affected correlation coefficient 0.3 which is small and positive. Correlation coefficient for healthcare index and recovered is 0.09 which is quite small to take in account. The overall value of populationtests-deathrate correlation coefficient is 0.2 which seems positive correlation but small to draw a conclusion.

\begin{table}[!h]
\centering
    \caption{\textbf{Correlation between COVID-19 Datasets and Different Enivironmental, Socio-Economic Factors}}     
    \label{tab:table}
    \resizebox{\textwidth}{!}{
    \begin{tabular}{|l|l|l|l|l|l|l|}
    \hline
    \multirow{2}{*}{Correlating factors} &
    \multicolumn{2}{|c|}{Pearson's coefficient} &
    \multicolumn{2}{|c|}{Spearman's coefficient} &
    \multicolumn{2}{c|}{Kendall's coefficient} \\
    & Range & Median & Range & Median & Range & Median \\
    \hline
    Temperature, Affected & -0.0019 to 0.6900 & 0.0866 & -0.0009 to 0.8023 & 0.0930 & -0.0050 to 0.6526 & 0.0777 \\
    \hline
    Humidity, Affected & -0.0093 to 0.5418 & -0.4951 & -0.0098 to 0.5189 & -0.4366 & -0.0018 to 0.4011 & -0.3155 \\
    \hline
    Pollution, Affected & - & -0.115 & - & -0.231 & - & -0.153 \\
    \hline
    Food security, Affected & - & 0.265 & - & 0.687 & - & 0.485 \\
    \hline
    Healthcare, recovered & - & 0.085 & - & 0.152 & - & 0.086 \\
    \hline
    Population tests, death rate & - & 0.157 & - & 0.101 & - & 0.073 \\
    \hline
    
  \end{tabular}
    }
\end{table}

\subsubsection{Correlation with Temperature}

In Table \ref{tab:tabtemp}, we have shown significant correlation values that have been found from the analysis. The values are significant as p-values are less than or equal to $\alpha$. Among 114 countries, values of 32 countries are significant to mention, 7 countries show negative correlation value and 25 countries show positive correlation value for Pearson's correlation analysis. Numbers are same for Spearman's and Kendall's also. Bangladesh, Belgium, Cuba, Iran, Japan, Mexico, Netherlands, USA, United Kingdom show values over 0.5 (positive). Colombia, Peru show values less than or equal to -0.5 (negative).

\begin{figure}[H]
    \centering
    \includegraphics[width=9cm]{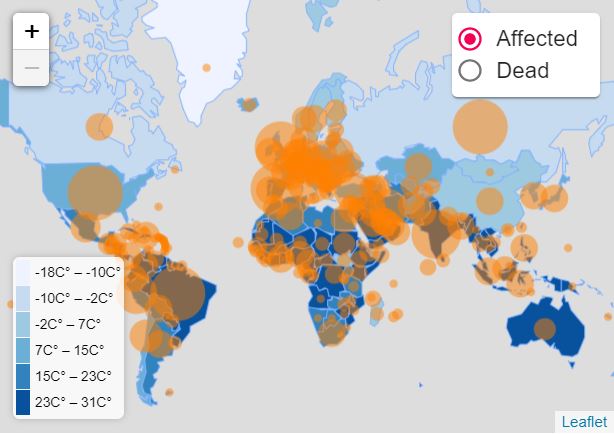}
    \caption{\textbf{Correlation between Temperature and Spreading of COVID-19}}
    \label{fig:tempMap}
\end{figure}

\begin{table}[!h]

    \caption{\textbf{Correlation between Temperature and Spreading of COVID-19}}
    \label{tab:tabtemp}
    \resizebox{\columnwidth}{!}{
    \csvreader[filter equal={\csvcoliv}{correlated},tabular=|l|l|l|l|l|l|l|c|,
    table head=\hline Country name & Pearson's correlation & P-value  & Spearsman's correlation & P-value  & Kendall's correlation & P-value  \\
    \hline,
    late after line=\\
    \hline]%
{temperaturelatonfindings.csv}{Country name=\Countryname,Pearson's correlation=\pearson,P-value of pearson's correlation=\pone,Spearsman's correlation=\spear,P-value of spearman's correlation=\ptwo,Kendall's correlation=\kenda,P-value of kendall's correlation=\ptree}%
{\Countryname & \pearson & \pone  & \spear & \ptwo  & \kenda & \ptree}%
    }

\end{table}

\subsubsection{Correlation with Humidity}

In Table \ref{tab:tabhumid}, we have shown significant correlation values that have been found from the analysis. The values are significant as p-values are less than or equal to $\alpha$. Among 114 countries, values of 30 countries are significant to mention, 24 countries show negative correlation and 6 countries show positive correlation value for Pearson's correlation analysis. Numbers are same for Spearman's and Kendall's, but the values slightly change. Andorra, Tunisia show values over 0.5 (positive). Belarus, Belgium, Denmark, Germany, Hungary, Lithuania, Netherlands, Panama, Russia show values less than or equal to -0.5 (negative).

\begin{figure}[H]
    \centering
    \includegraphics[width=9cm]{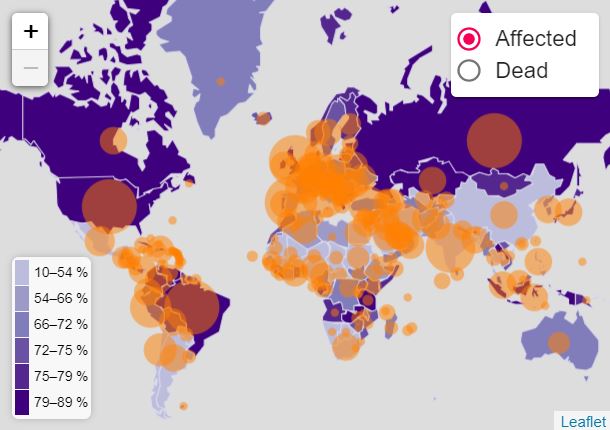}
    \caption{\textbf{Correlation between Humidity and Spreading of COVID-19}}
    \label{fig:humidMap}
\end{figure}

\begin{table}[!h]

    \caption{\textbf{Correlation between Humidity and spreading of COVID-19}}
    \label{tab:tabhumid}
    \resizebox{\columnwidth}{!}{
     \csvreader[filter equal={\csvcoliv}{correlated},tabular=|l|l|l|l|l|l|l|c|,
    table head=\hline Country name & Pearson's correlation & P-value  & Spearsman's correlation & P-value  & Kendall's correlation & P-value  \\
    \hline,
    late after line=\\
    \hline]%
    {humidityrelatonfindings.csv}{Country name=\Countrynamet,Pearson's correlation=\pearsont,P-value of pearson's correlation=\ponet,Spearsman's correlation=\speart,P-value of spearman's correlation=\ptwot,Kendall's correlation=\kendat,P-value of kendall's correlation=\ptreet}%
{\Countrynamet & \pearsont & \ponet  & \speart & \ptwot  & \kendat & \ptreet}%
    }

\end{table}

\subsubsection{Correlation with Pollution}

According to the collected data from \cite{numbeo}, we have analyzed the countries having the pollution ranking from worst to best and total number of affected cases so that, we can understand how countries are affected by COVID-19 if it is polluted enough. Some countries having good ranking (index between 10 and 30) such as, Germany, Canada, Netherlands are large in numbers of affected cases. On range 30 to 50, United States, Spain, France, United Kingdom are devastating in numbers of affected cases. Badly polluted countries such as, Peru, India, China, Bangladesh, Iran are having a notable rise in number of affected cases.

\subsubsection{Correlation with Healthcare}

According to the collected data from \cite{numbeo}, we are mentioning some of our observations to understand how countries are recovered from COVID-19 if it is well-equipped in health system. Some good ranked countries (index between 87 and 65) such as, South Korea, Spain, Germany, United States, Denmark, France, Italy are good in numbers of recovered cases. In spite of being in top ranked list, United Kingdom, Norway, Australia dissatisfy us in numbers of recovered cases. On range 60 to 50, Russia, Brazil, Peru, Poland are good in numbers of recovered cases. Badly indexed countries such as, Bangladesh, Egypt are not so satisfactory in numbers.

\subsubsection{Correlation with Food Security}

According to data from \cite{food}, we can see ranking of food security from good to bad and total number of affected cases. USA, United Kingdom, Brazil, Italy are good both in food security and COVID-19 cases which is alarming fact to observe. Comparing with food security ranking, we can observe, many average and low ranked countries are also low in total affected case numbers.
    
\subsubsection{Correlation with Population Tests}

We have collected population tests data from \cite{death}. Counting on population tests is important because symptomatic mass tests suppress the disclosure of appropriate COVID-19 cases. If total population test is increased, it will reveal more cases. Death rate is dependent on total cases also. As a result, finding correlation among population tests and death rate is important. USA has performed total 16757346 tests till May 29, 2020 and has death rate of 6. Although Russia has performed 10000061 tests, the death rate is 1.2. Italy, France, Belgium, Netherlans, Maxico show high mortality rate though performed tests are less in numbers compared to USA, Russia. UK has scored 14 in mortality rate and has performed 67853964 tests. In a nutshell, acceptable population tests dependency can not be noticed from the above observations.

Temperature, humidity data give us an overview of geographic position of a country and correlation of them with COVID-19 lead us to geographic investigation and responses to the pandemic. How quickly the virus can be spread and how to reduce transmission within communities can be answered through this kind of analysis.

Correlation findings with food security and pollution index lead us to answer the questions how individual country is vulnerable to the pandemic with respect to cost of living of corresponding country. It will help to draw difference between first world and third world and their plans of taking measures to mitigate the pandemic will be different. 

Analysis on healthcare index and population tests measurements lead to focus more on maintaining healthcare services of a country. The analysis will help to minimize disruptions caused by necessary community health interventions and support health planning and the allocation of appropriate resources within the healthcare system.

\subsection{Future Prediction}

We ran our mathematical model of future prediction on different countries by collecting affected, deaths, recovered, and demographic data \cite{popPyra} \cite{statista} from various sources. At first, we determined the underlying constants of the compartmental model (\textbeta, \textlambda$_d$, \textlambda$_r$). We ran the basic SIRD model on the existing historical data of the corresponding country to determine the constants. We tweaked the lockdown and social distancing indicating constants (\textdelta${_1}$, \textdelta${_2}$, \textdelta${_3}$, \textdelta${_4}$) to get three different states: \emph{Lockdown Imposed and Social Distancing Maintained}, \emph{Lockdown Released and Social Distancing Maintained}, and \emph{Lockdown Released and Social Distancing Discarded.}\\

Therefore, after determining all the constants, we predicted the number of affected and the number of deaths for the next 60 consecutive days from the latest data. The latest data we used is of May 29. Here we present our results by running the model on Bangladesh, India, Russia, United States, Canada, Chile, Mexico, and Brazil. We specifically picked these countries to cover various aspects. Bangladesh, India, Chile, Mexico, and Brazil are experiencing an increasing number of affected per day, showing a nearly exponential trend on the graph. On the other hand, Russia has seen a lot of confirmed cases. However, the slope of the curve is still giving a positive trend. Canada has been able to flatten their curve for now, but still the infection is spreading to some extent. Finally, United States has seen the highest number of confirmed cases and the infection is still spreading a lot at the country. Besides, all countries are heavily populated except Canada and Chile. These countries convey a good demographic significance. All these eight countries represent four clusters altogether, based on the causes of death and health risk factors \cite{Isikhan:2018}.

\subsubsection{Prediction of Affected Numbers}
    \begin{figure}[p]
        \centering
        \subfloat[\emph{\textbf{Bangladesh}}]{{\includegraphics[width=\w\textwidth , height=\h\textheight]{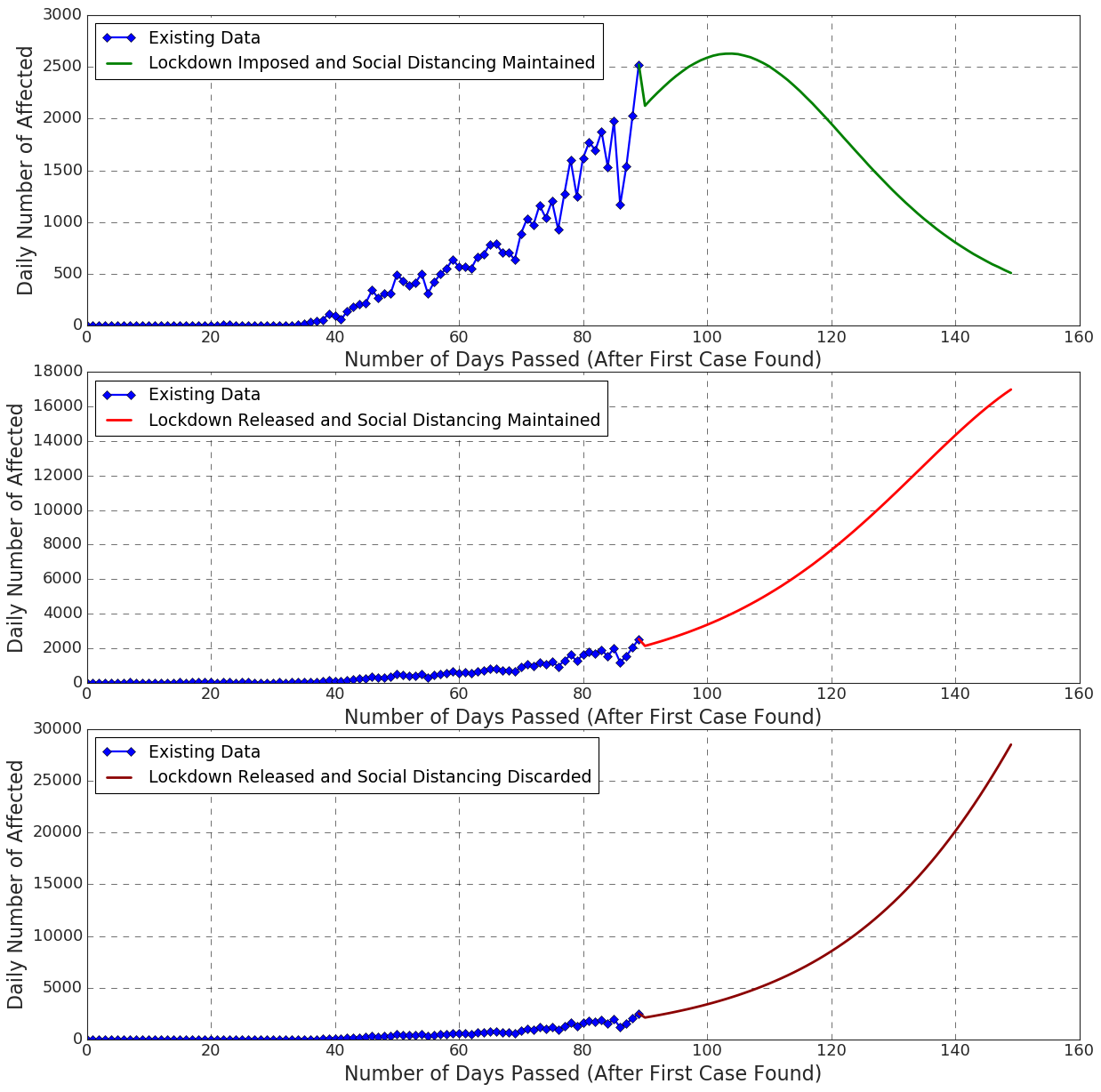} }}%
        \quad
        \subfloat[\emph{\textbf{India}}]{{\includegraphics[width=\w\textwidth , height=\h\textheight]{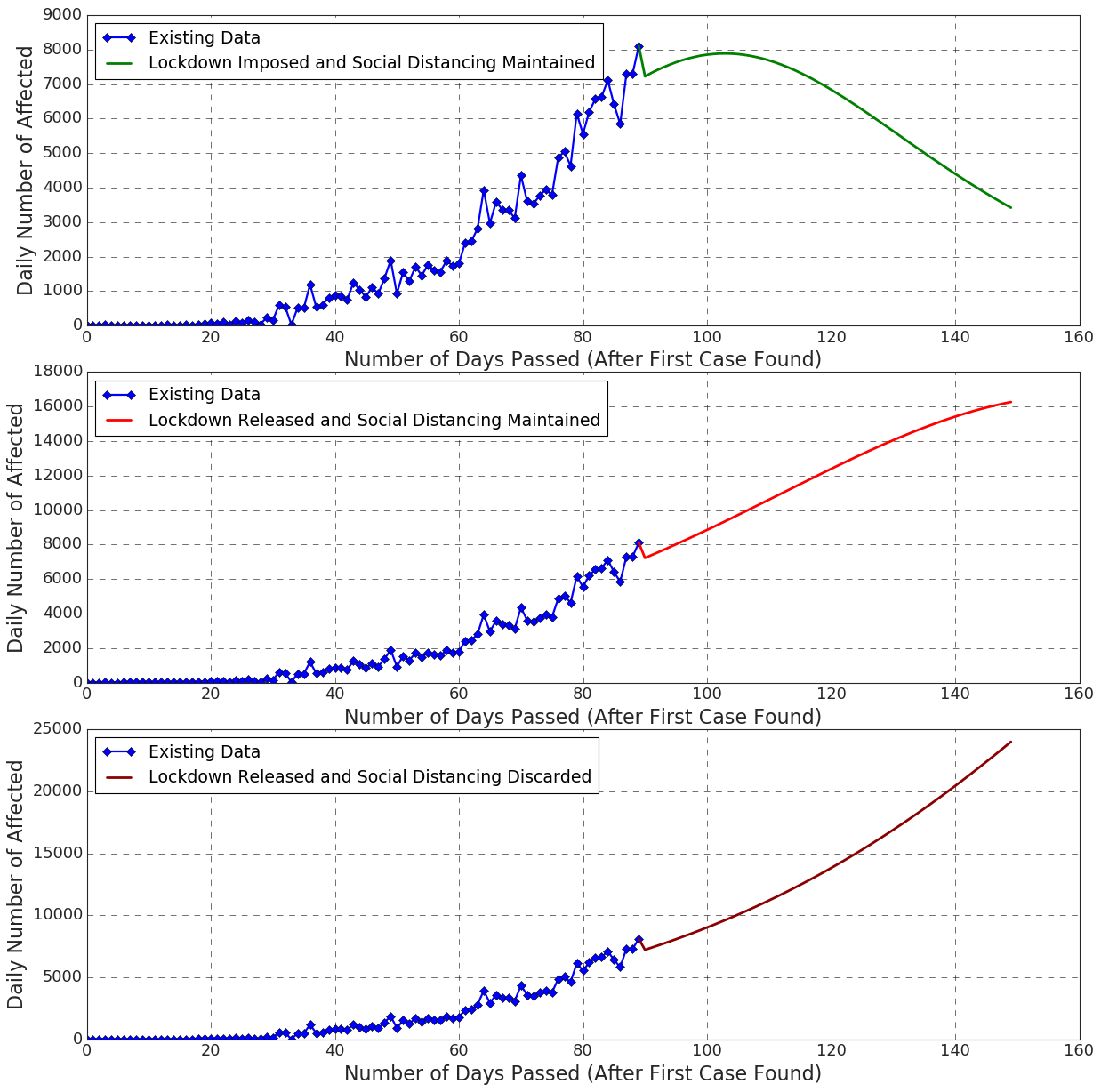} }}%
        \quad
        \subfloat[\emph{\textbf{Russia}}]{{\includegraphics[width=\w\textwidth , height=\h\textheight]{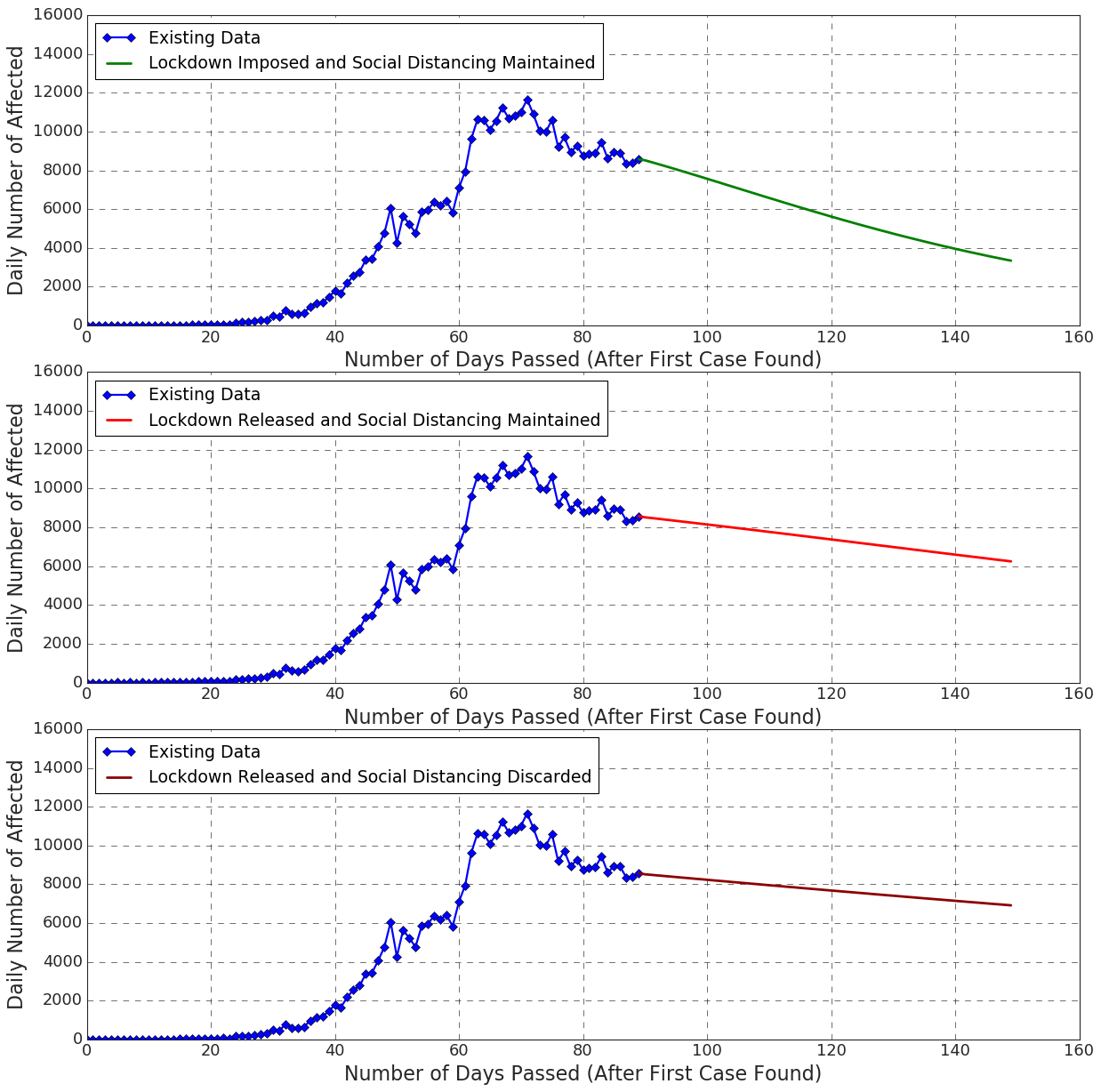}}}%
        \quad
        \subfloat[\emph{\textbf{United States}}]{{\includegraphics[width=\w\textwidth , height=\h\textheight]{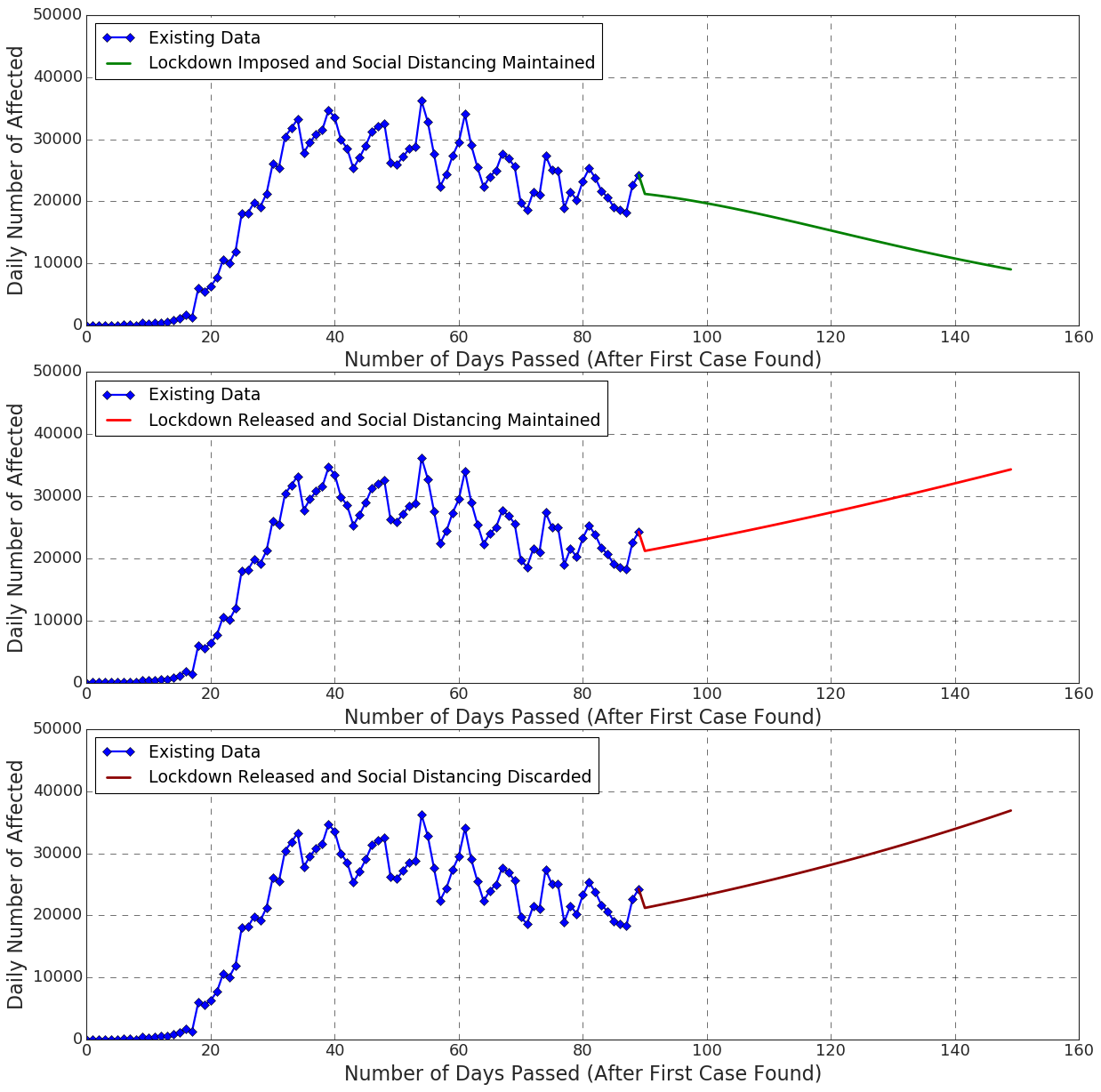}}}%
        \caption{\textbf{Prediction of Daily Number of Affected for 60 days from Latest Data(1)}}%
        \label{a:h1}%
    \end{figure}
    
    \begin{figure}[p]
        \centering
        \subfloat[\emph{\textbf{Canada}}]{{\includegraphics[width=\w\textwidth , height=\h\textheight]{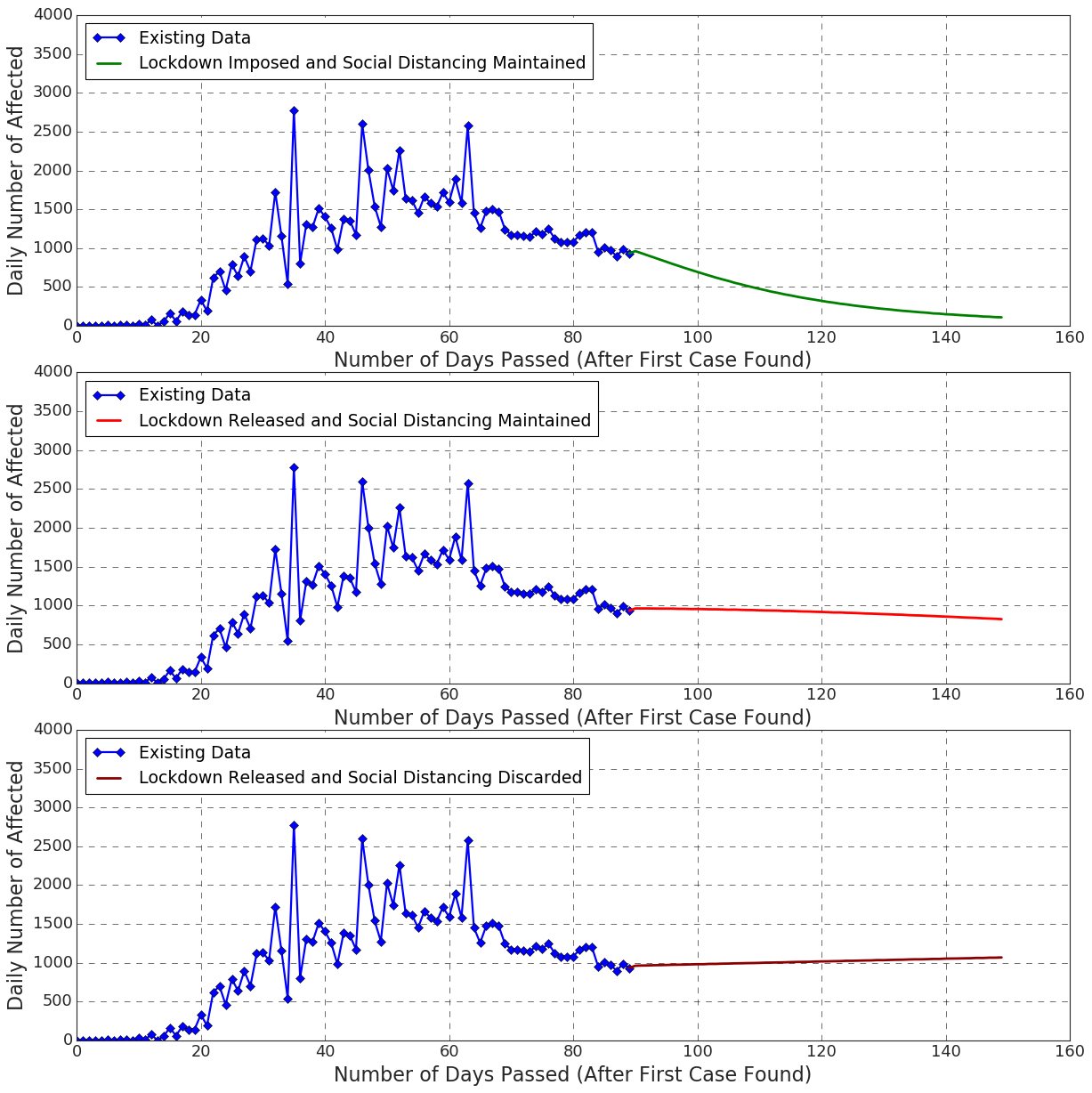} }}%
        \quad
        \subfloat[\emph{\textbf{Chile}}]{{\includegraphics[width=\w\textwidth , height=\h\textheight]{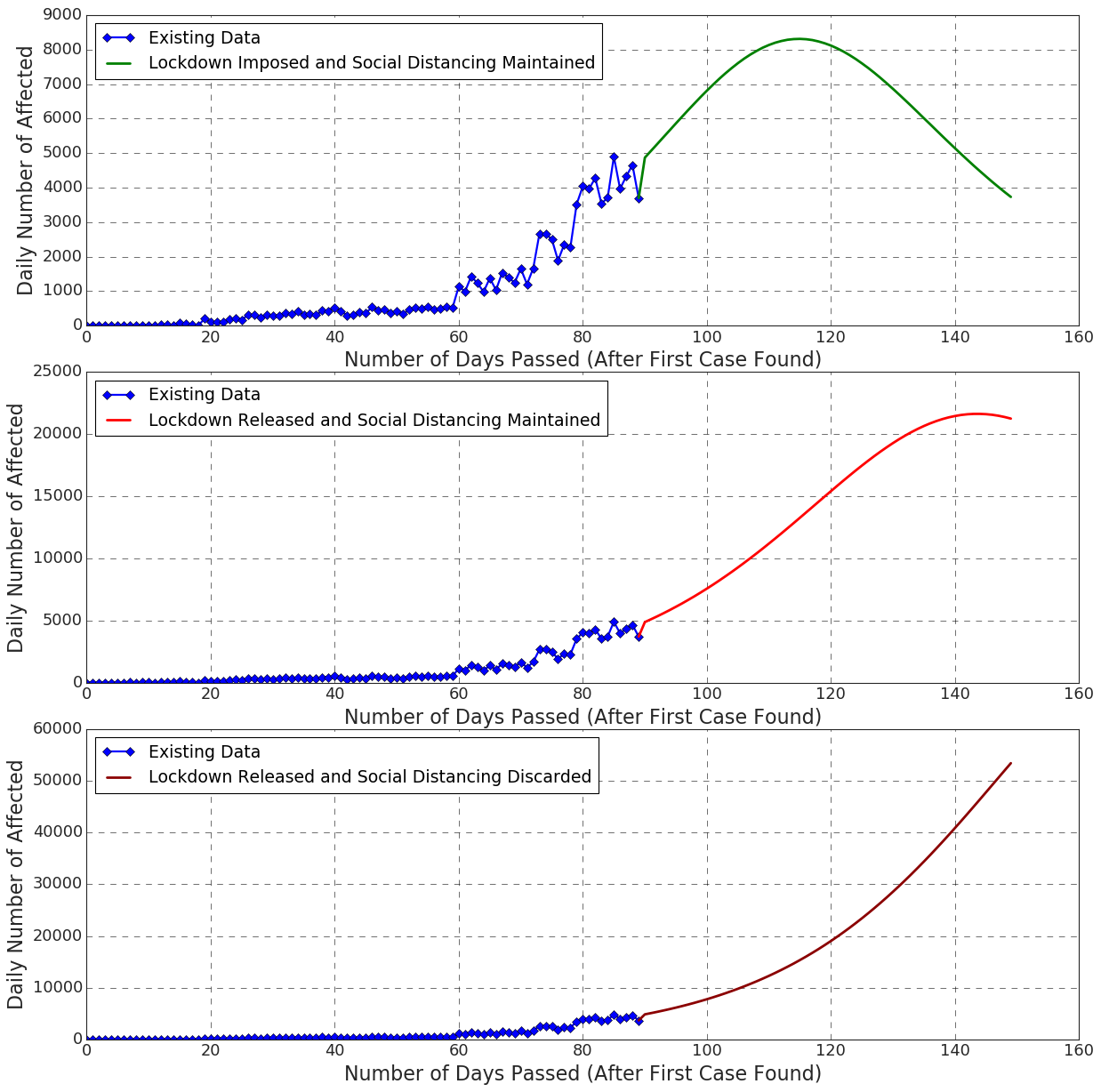} }}%
        \quad
        \subfloat[\emph{\textbf{Mexico}}]{{\includegraphics[width=\w\textwidth , height=\h\textheight]{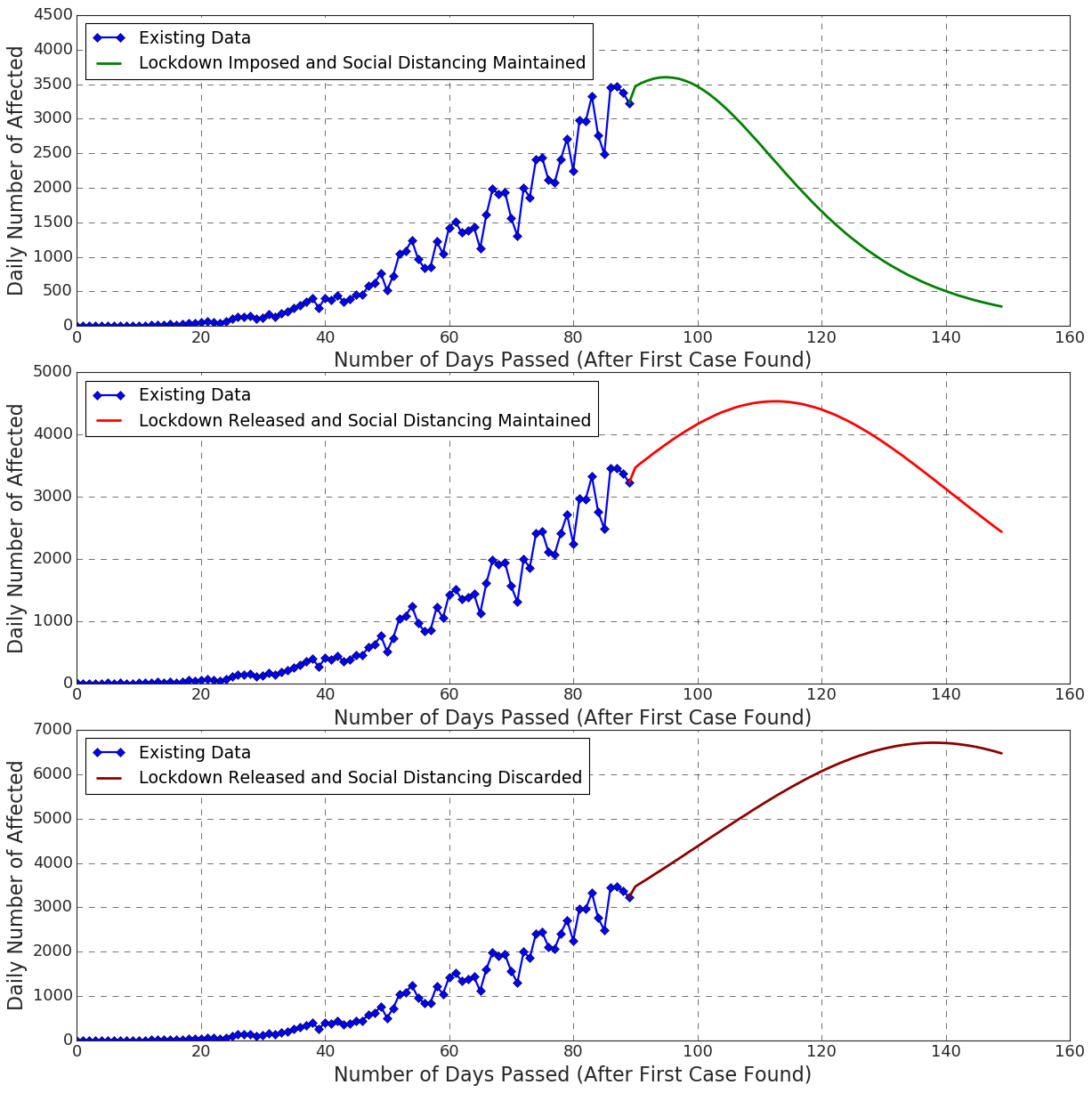}}}%
        \quad
        \subfloat[\emph{\textbf{Brazil}}]{{\includegraphics[width=\w\textwidth , height=\h\textheight]{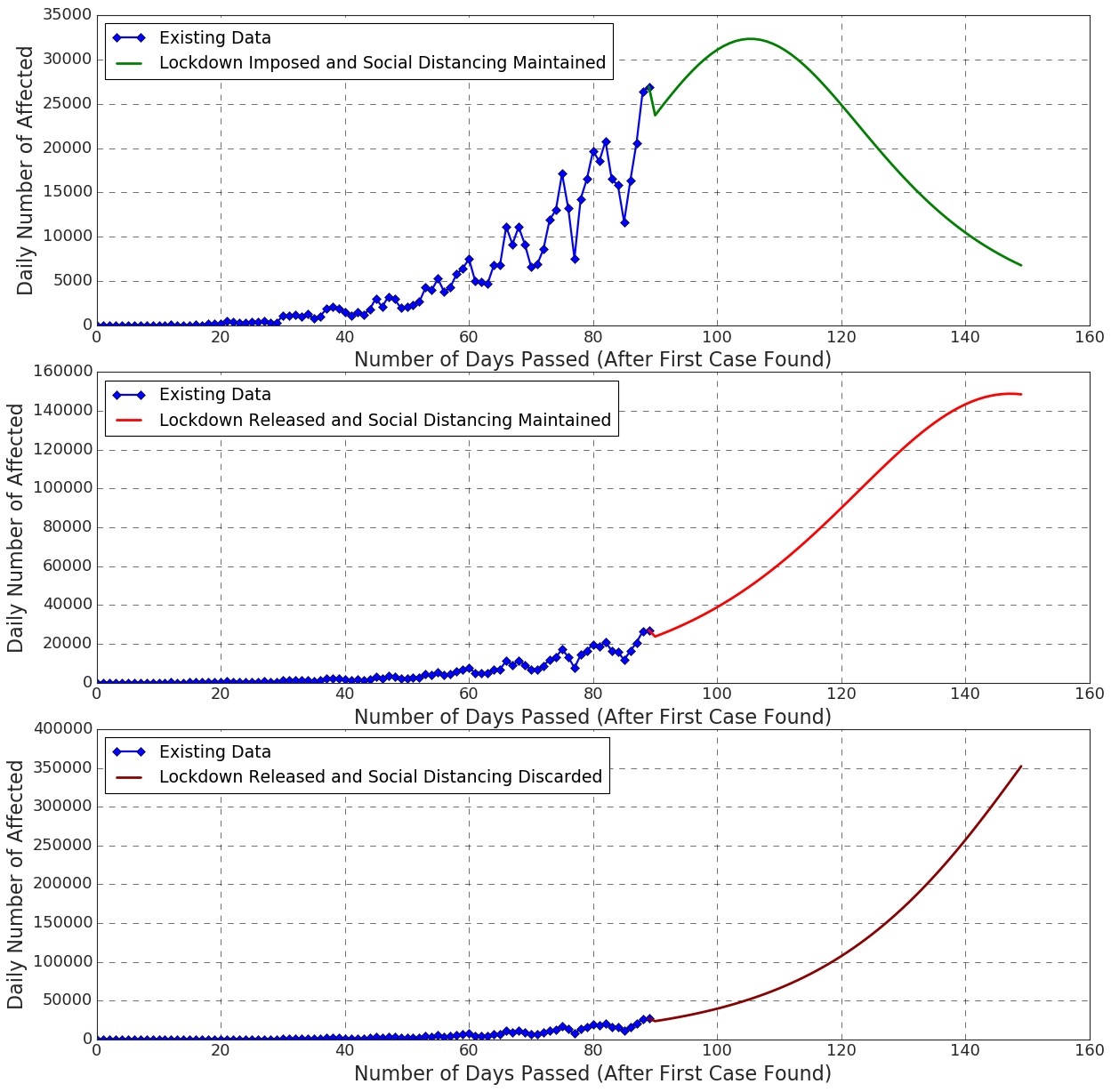}}}%
        \caption{\textbf{Prediction of Daily Number of Affected for 60 days from Latest Data(2)}}%
        \label{a:h2}%
    \end{figure}
    
    We have found that we can 'flatten the curve' of number of infected individuals per day within a month in Bangladesh, India, Russia, Mexico, Chile, and Brazil (Figure: \ref{a:h1}, \ref{a:h2}), if we follow the social distancing measures and the lock-down restrictions. Since United State's and Canada's curves have already flattened, our findings say the number of infections will be reduced to some extent in these countries ((Figure: \ref{a:h1}, \ref{a:h2}). \newline
    However, the curve shows a pure exponential trend for Bangladesh, India, Russia, and Brazil (Figure: \ref{a:h1}, \ref{a:h2}), if the lock-down is released and the social distancing is discarded. \newline
    We also measured the impact of a released lock-down with a maintained social distancing. Our finding is, this measure will not help much. Although in this case, the curve is less sloppier than that of discarding social distancing one. Yet, the curve does not become flatten within \emph{60 days}. 

\subsubsection{Prediction of Deaths Numbers}

        \begin{figure}[p]
            \centering
            \subfloat[\textbf{Bangladesh}]{{\includegraphics[width=\w\textwidth , height=\h\textheight]{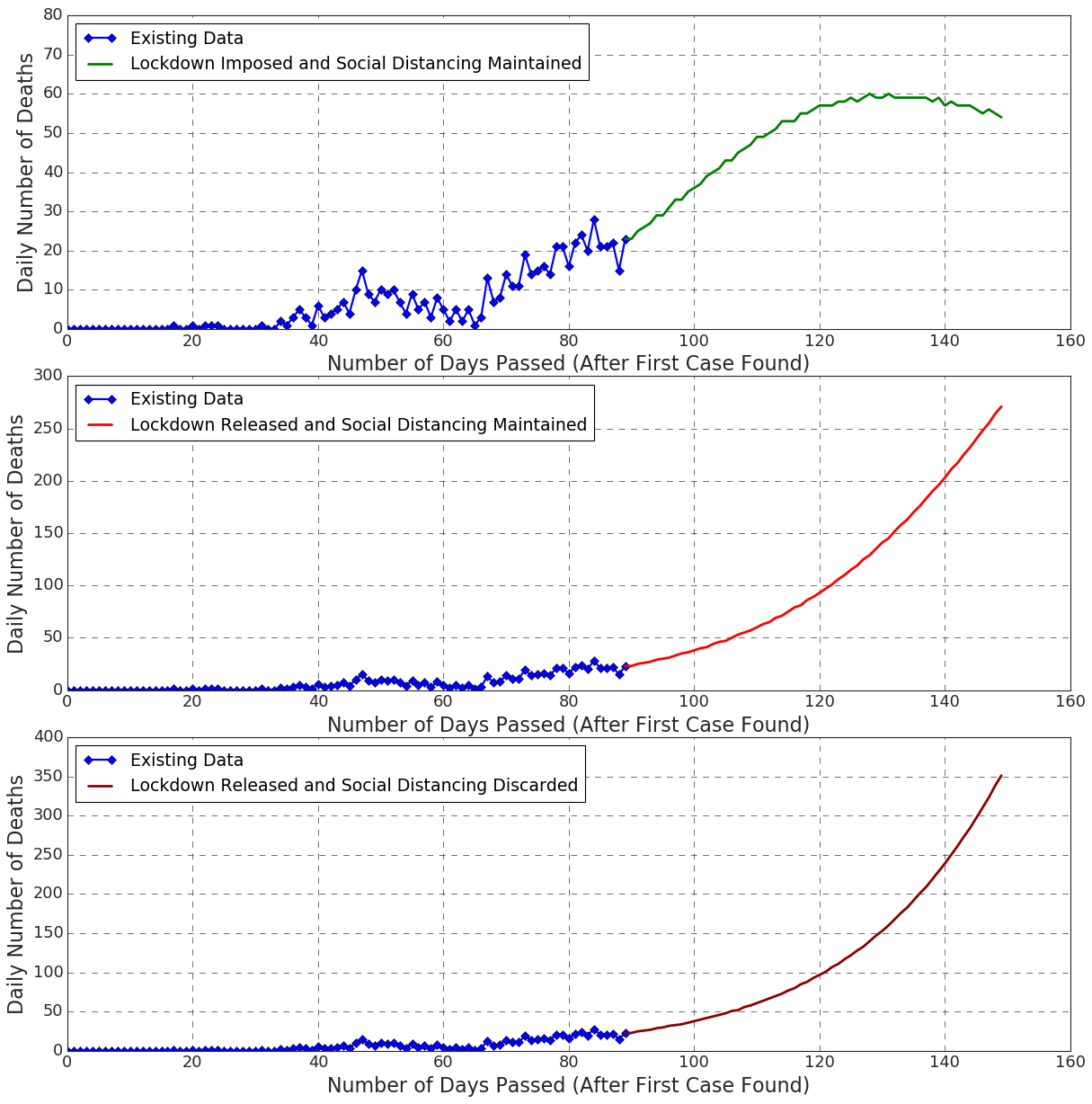} }}%
            \quad
            \subfloat[\textbf{India}]{{\includegraphics[width=\w\textwidth , height=\h\textheight]{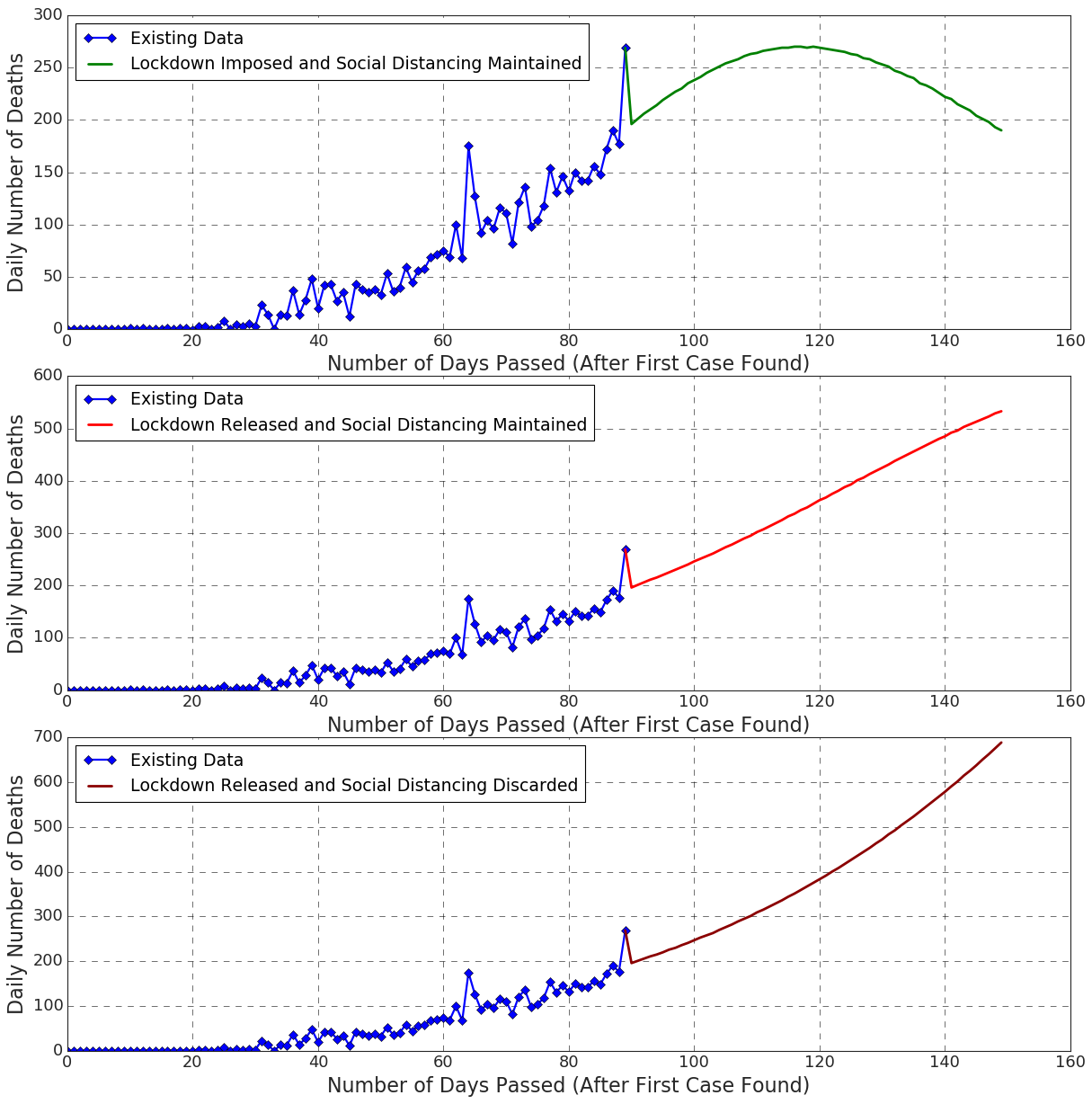} }}%
            \quad
            \subfloat[\textbf{Russia}]{{\includegraphics[width=\w\textwidth , height=\h\textheight]{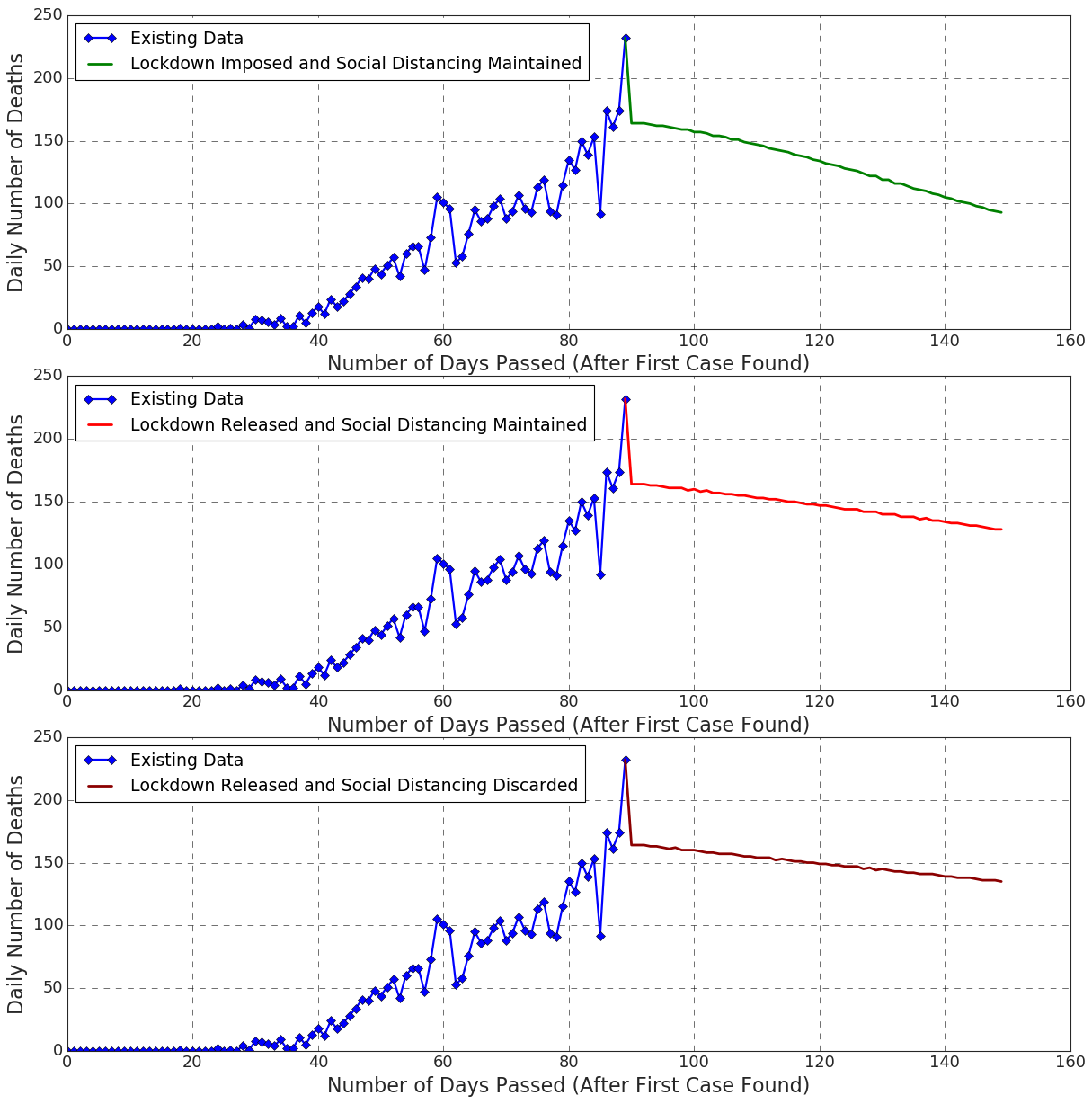}}}%
            \quad
            \subfloat[\textbf{United States}]{{\includegraphics[width=\w\textwidth , height=\h\textheight]{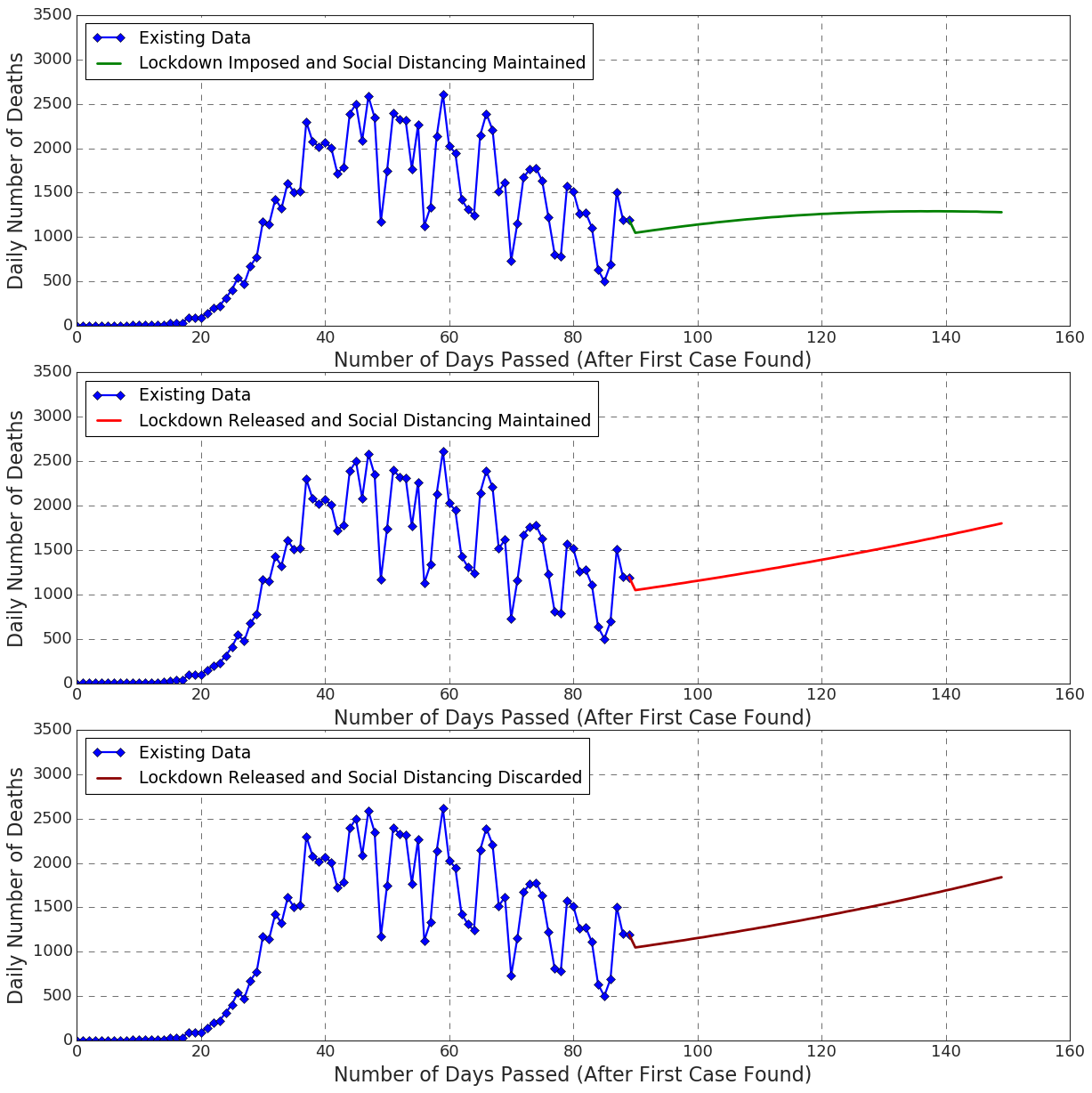}}}%
            \caption{\textbf{Prediction of Daily Number of Deaths for 60 Days from Latest Data(1)}}%
            \label{d:h1}%
        \end{figure}
        
        \begin{figure}[p]
            \centering
            \subfloat[\textbf{Canada}]{{\includegraphics[width=\w\textwidth , height=\h\textheight]{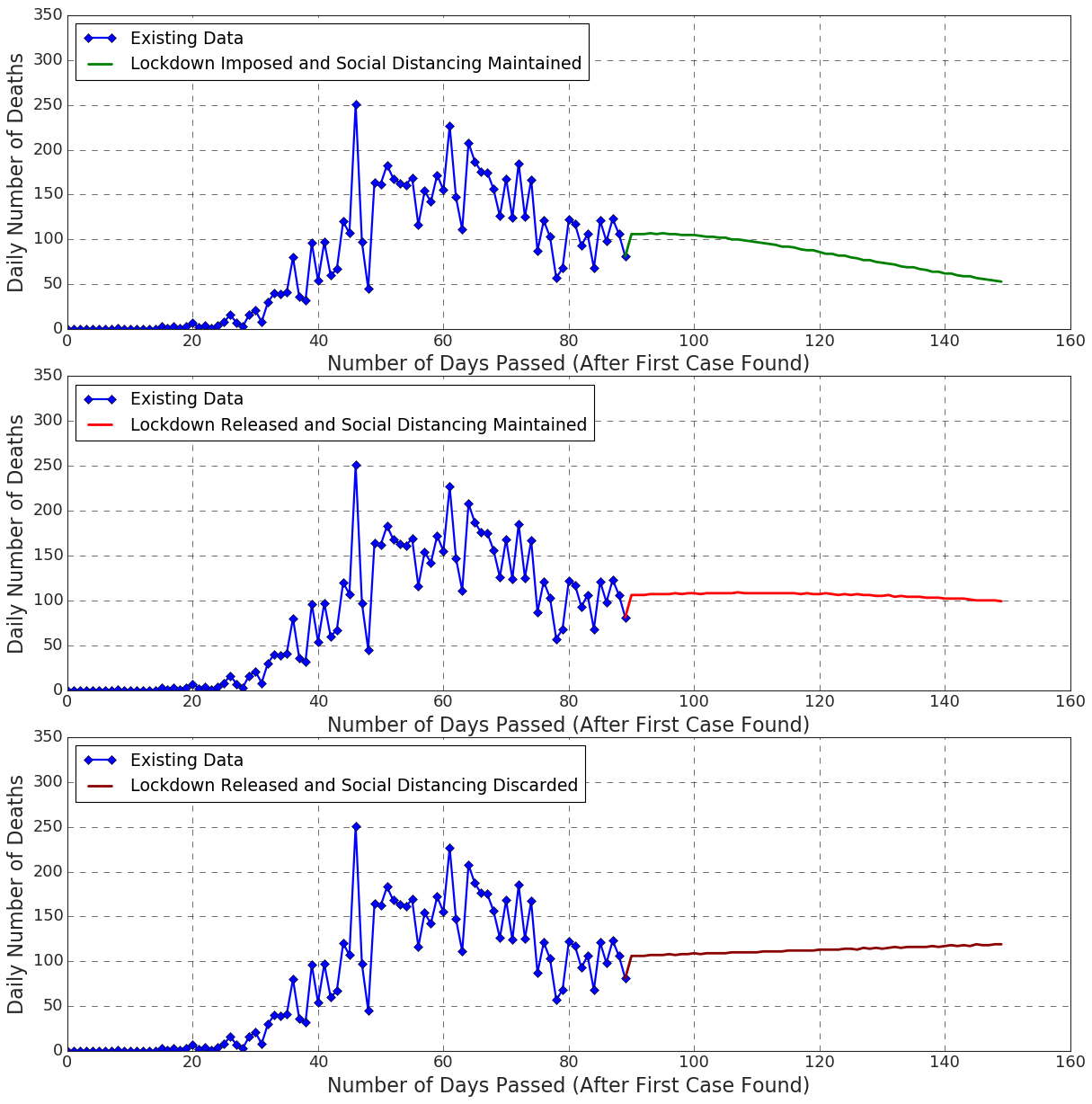} }}%
            \quad
            \subfloat[\textbf{Chile}]{{\includegraphics[width=\w\textwidth , height=\h\textheight]{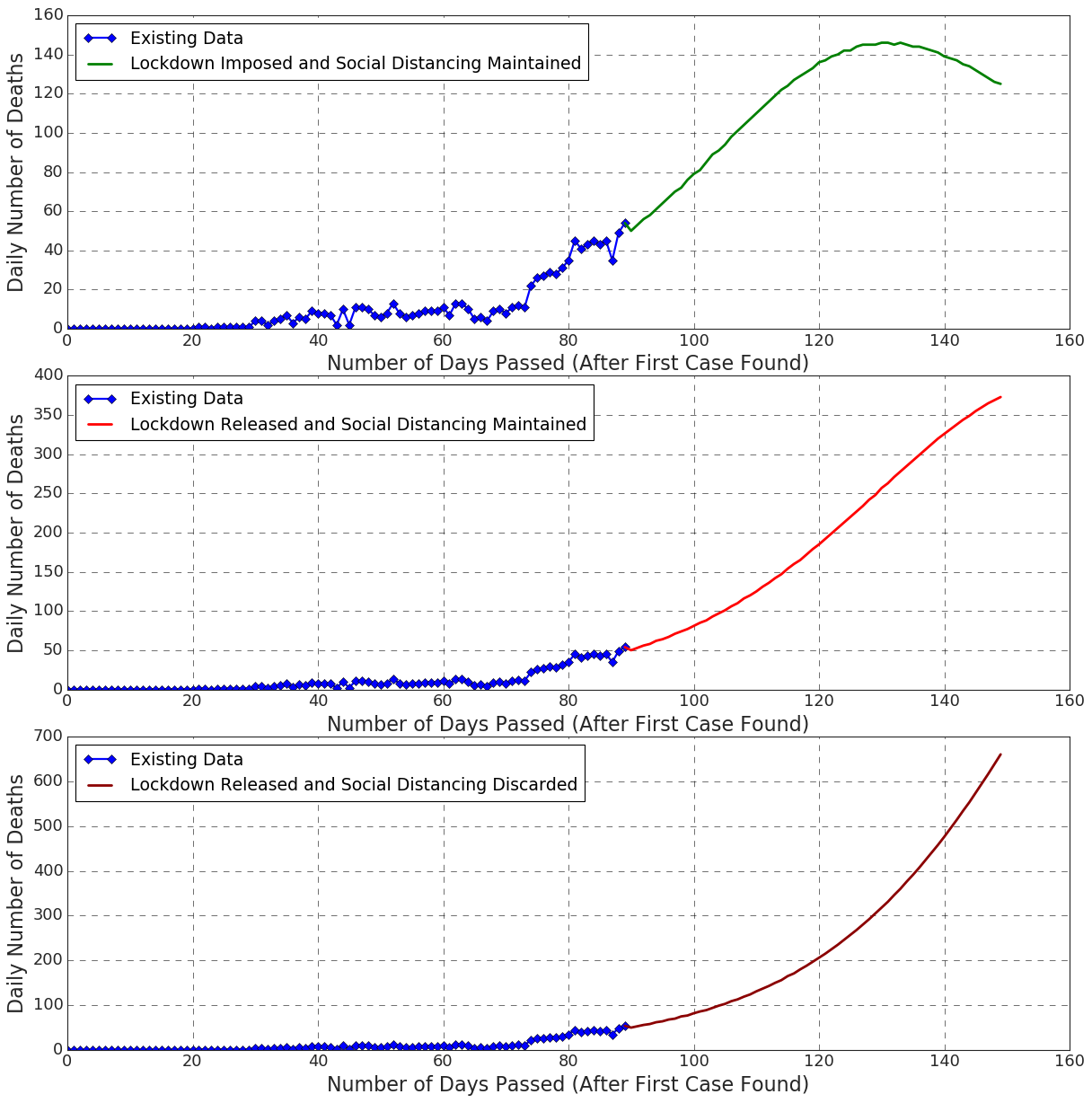} }}%
            \quad
            \subfloat[\textbf{Mexico}]{{\includegraphics[width=\w\textwidth , height=\h\textheight]{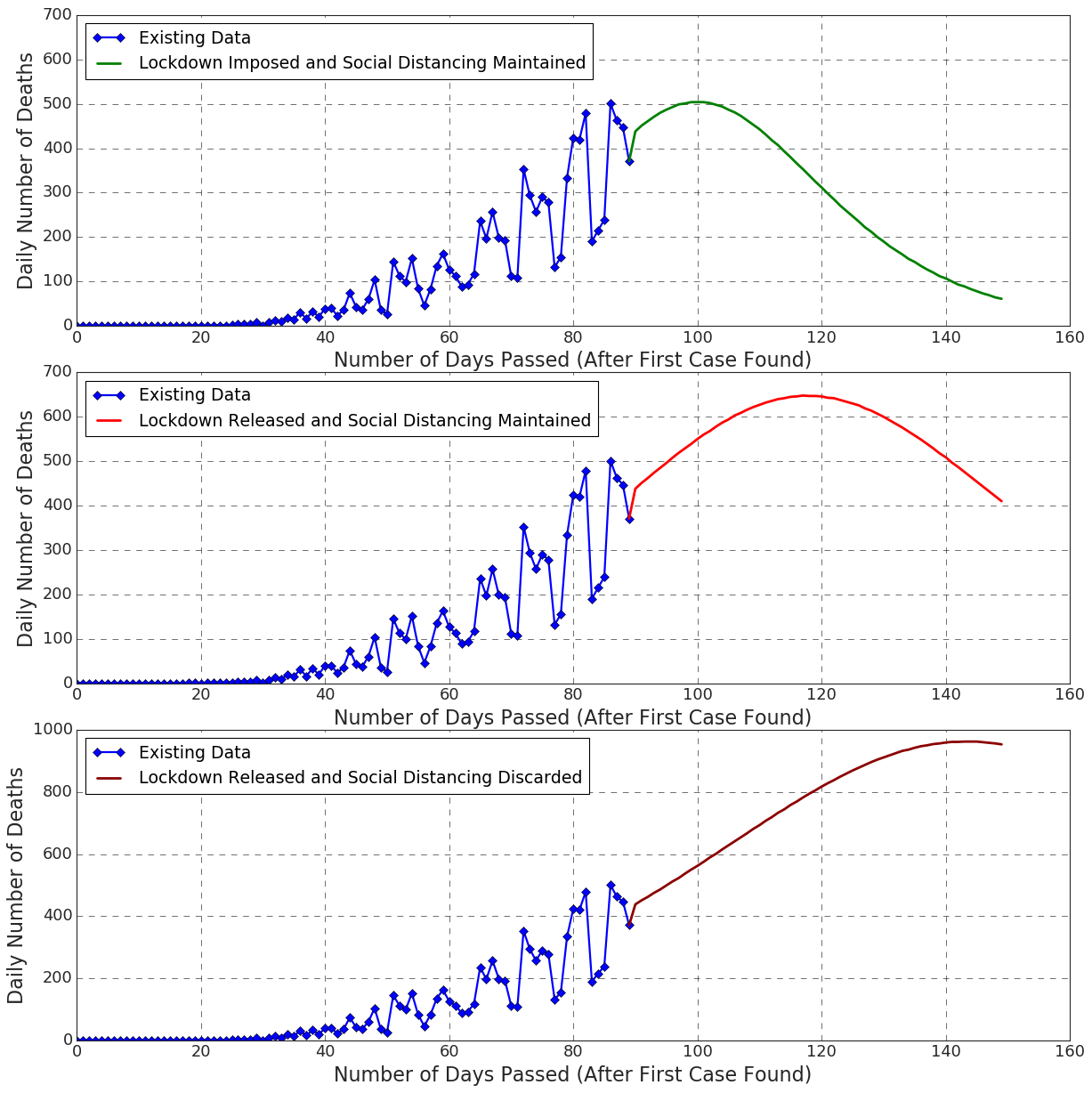}}}%
            \quad
            \subfloat[\textbf{Brazil}]{{\includegraphics[width=\w\textwidth , height=\h\textheight]{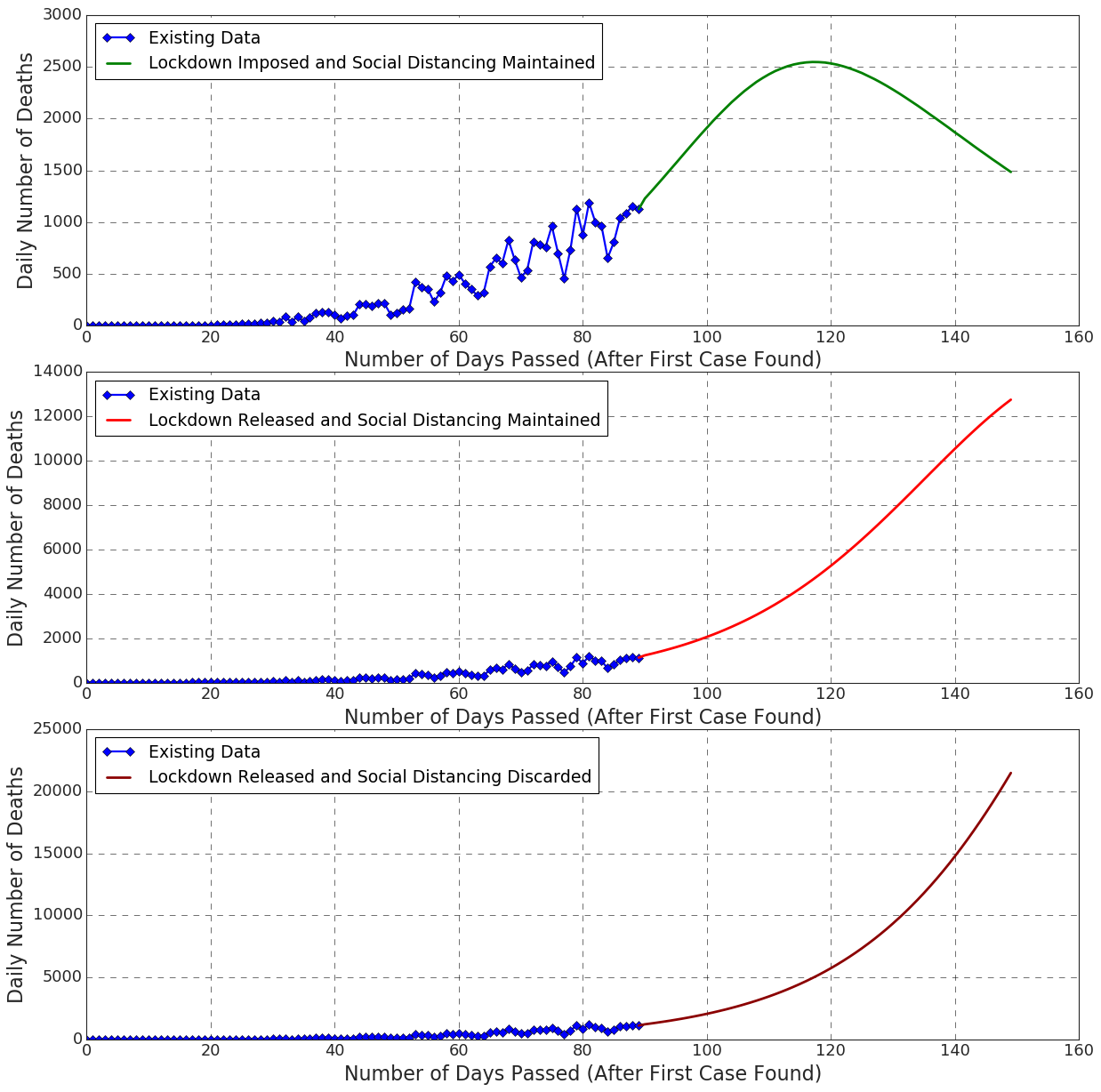}}}%
            \caption{\textbf{Prediction of Daily Number of Deaths for 60 days from Latest Data(2)}}%
            \label{d:h2}%
        \end{figure}
        
         We have found that the number of deaths per day follows the pattern of the number of infected individuals per day (Figure: \ref{d:h1}, \ref{d:h2}). Like before, we have found that we can limit the total number of deaths to a great extent, if we follow the social distancing measures and the lockdown restrictions. 
        However, if the lockdown is released, we can find that both the curves of social distancing measured and social distancing discarded follow the same trend (Figure: \ref{d:h1}, \ref{d:h1}). \\
        Vital dynamics contribute to the number of susceptible with \textchi  $=$  $($\textmu $-$ \textGamma$)$$\times$$N$ number of people. Since \textchi~ is mostly positive \cite{bdrate}, it only adds \textchi number of people to the susceptible compartment. However, it still remains a research question \cite{babycorona} whether newborn babies not inheriting the COVID-19 virus from the mother are actually susceptible to the virus or not. Besides, while we are considering vital dynamics, we are considering the numbers of accident deaths that would happen in a normal continued lifestyle. It might compensate the numbers of unreported deaths due to COVID-19 virus \cite{ZhaoUnreported}.\newline
        The reinfection fact that we considered while building the model, is very insignificant in the results. The reinfection is much less in the case of COVID-19 \cite{Victor:2020}. If it were significant enough, we would get a clear oscillating trend in the graphs. However, if the reinfection rate increases in the future due to the mutation of the virus \cite{ZhaoMutation}, the reinfection numbers may become significant.
\section{How do our analysis facilitate policy making} 
Based on outcomes of correlations and future predictions of the pandemic (RQ1 and RQ2), we now elaborate how the insights show path to agile policy making during pandemics like COVID-19 (RQ4).

\subsection{Mitigating Community Transmission of Disease}
Continuous disease surveillance on different aspects of a country and predictions of numbers of affected and deaths give a strong overview of repercussion of the pandemic on a country's public health and cost of living. Understanding these impacts elaborately will help the policy makers to take useful measures to mitigate the spread of the disease among the community. For example, rising cases of COVID-19 in Bangladesh, Brazil, India indicates that more strict lockdown and social distancing should be imposed to mitigate transmission of the disease locally. Government agencies and healthcare institutions can get ideas of recurring events that contribute to outbreak of the disease. As gatherings contribute in spreading of disease vastly, restrictions should be imposed strictly. Policy makers should try their best to keep the community transmission of the disease as low as possible \cite{community}.

However, this is easier said than done. There are critical ramifications to imposing lock-downs. These include economic losses to local populations, lack of schooling opportunities (and most worryingly, for children with special needs), victims of domestic assault not being to able to seek help, and so much more. We believe that mining social media at population scale can provide valuable insights on unknown perceptions and hardships of the population for governmental agencies to do smarter lock down measures.  We are aware of many recent studies in this space [cite], and in-times of lockdown with users spending more time on the Internet, the scale of data mined will only massively increase and can be put to good use.

\subsection{The Importance of Contact Tracing} In combating communicable diseases (especially HIV, Pertussis and sexually transmittable diseases), contact tracing has proved to be a vital asset. However, in the case of COVID-19, the sheer speed and scale of virus spread meant that traditional methods of manual contact tracing by a few public health workers were going to be insufficient. Across the world, digital methods of contact tracing using smartphones was seen as a viable option. Many governments, private companies and universities developed a slew of contact tracing apps (each of which centered on using short range Bluetooth like communication signals) to record proximity to other devices via mutually shared secrets. When a user reports that he/she was infected with Covid-19, notification to those that could have been exposed was enabled by these systems, based on secrets shared. 

Unfortunately, despite privacy and security guarantees, adoption was very poor. If digital contact tracing methods could have been adopted earlier, it is likely that infection rates could have slowed down. The main reason for lack of adoption was that common citizens simply got {\em scared} that the governments could {\em track} them. Allaying such fears using carefully planned mechanisms should be attempted now, so that public trust can be earned, and such systems may potentially help combat future spread of COVID-19, and possibly any future diseases like this.

Furthermore, our analysis of historical data can identify populations and area of high risk and how infections are emerging in different geographical locations. This information, when coupled with contact tracing mechanisms and other useful datasets (e.g., population test, pollution, healthcare system, temperature, humidity) can help government agencies identify areas that are under imminent risk, requiring immediate attention. This can also help the government prioritize the allotment of its limited resources to different sectors based on the limitations of the sectors.

\subsection{Adopting Necessary Health Planning Measures}

Apart from controlling and mitigating the deadly impact of the pandemic, allocation of appropriate resources within healthcare system is equally important. Correlating healthcare index with COVID-19 cases reflects how badly recovery rate and death rate can be affected due to bad healthcare system of a country. Total number of population tests address how public health system is tracking the epidemic. If enough tests are not being carried out, the risk of undetected community transmission and resurgence increases. Also number of detected deaths due to COVID-19 will not come into light. Government officials and policymakers get a grasp of on the ground realities of public health system. Public health officials can use this indicators to prepare public health system more resourceful to mitigate the deadly effects. For instance, healthcare systems of Bangladesh \cite{bdh}, Egypt \cite{egh}, Brazil \cite{brzh}may lead to increasing cases of deaths. Thus, the policy makers should focus on necessary health plans to take over the situation.

\subsection{Encouraging Situational Awareness}
How maintaining social distancing and imposing lockdown can help reduce the spreading of virus? We expounded on this question and found out it helps all the local community. Furthermore, we conducted our study in a country basis modular way. It is also evident that traveling to another country in the ongoing circumstances could increase the traveller's susceptibility to the disease or the countryman's, whom he is visiting to. This highlights the importance of effective non-pharmaceutical public health interventions. Policy makers can use mass media in this regard.  During this time of the pandemic, mass media played an indispensable role in promoting awareness and solidarity to people around the world \cite{media}. Limiting interaction to a few repeated contacts, imposing community quarantine etc. can be implemented by individuals and policy makers as well. There are some other intriguing effects of the lockdown as well. Study shows that crime has also declined in a significant way in this lockdown \cite{crime}. Study also shows that there was a notable reduction in air pollution which is good for environment and people health \cite{polldown}.

\section{Discussion}

Our project consisted of accumulating a lot of data from a variety number of sources, representing the data visualizations in an interactive way, finding fascinating results upon analyzing the data, and creating a forecasting model with the help of existing data. We tried to determine whether there is any correlation between affected cases and environmental factors, recovered cases and healthcare system, affected cases and food security, population tests and death rate. We perceived a significant amount of positive correlation in many countries of different continents. But the reason of the positive correlation is not clear enough to draw a conclusion, though temperature has shown good amount of correlation for some countries. Median humidity-affected correlation is negative. So weather and geographical position may have less impact on the outbreak of COVID-19. We need more geographical and weather data to find out whether environmental factors influence spreading of COVID-19 or not. Besides, we also have found that high ranked countries in healthcare system and food security are on the front line in combating this pandemic. On the other hand, as per our prediction model findings, we should strictly maintain the social distancing measures and lockdown restrictions. It will be hard to combat the incoming fluxes of new infections and deaths if the lockdown is released. These incoming fluxes may overflow the medical capacity of the countries and make the situation much worse. We also observed that the rate of reinfection is negligible in the cases of COVID-19.

\section{Scope For Further Research}

Our project was completely data based. But due to insufficiency of some crucial data, we could not make our work more substantial. We were in great dearth of sophisticated hospitalization data, lockdown schemes and schedule data, asymptomatic patient data, data of innate immunity against COVID-19, etc. We could make our prediction model more robust using these data.\\
When this pandemic ends, we would perform more in-depth analysis on the existing data to uncover important findings. These findings might be useful to the researchers, scientists, doctors, and relevant stakeholders in the outbreak of a pandemic like this in future. Besides, we are looking forward to work on the impacts of the pandemic. We would thus collect, analyze, and visualize the various socio-economic and demographical data that might be altered significantly by the pandemic. Upon accumulating enough data, we would also address the psychological effects of the pandemic on susceptible individuals, infected individuals, individuals closely related to the infected ones, and recovered individuals.

\section{Conclusion}


Policy making may be considered as the most important aspect during the ongoing COVID-19 pandemic, as it circumscribes the overall functionality of a region during and after the pandemic. 
An important basis of effective policy making is insightful data, which is even more important in case of a pandemic due to its spanning over the spatio-temporal domain. Accordingly, in this work, we focus on how to provide authentic and reliable comprehensive data to the policy makers. To do so, we explored correlations between the pandemic spreading and a number of different socio-economic and environmental factors ranging from food security index to temperature and humidity. We covered different aspects related to spreading, recovery, and death owing to the pandemic in its consideration. Besides, we demonstrated a robust future prediction of the pandemic over different regions of the world in different time frames. We presented outcomes of all the work in an integrated manner in our dashboard. Finally, we elaborated how the outcomes can influence efficient policy making from diversified contexts.

\begin{acks}

We would like to acknowledge (\href{https://retail.ai/}{Intelligent Machines Limited}) for giving us privilege to deploy our website on AWS. We also acknowledge (\href{https://www.heroku.com/}{Heroku}) for hosting version 1.0 of CORONOSIS. Finally, We acknowledge (\href{https://www.numbeo.com/cost-of-living/}{Numbeo}) for giving us access to their pollution index and healthcare index data. 
\end{acks}

\bibliographystyle{ACM-Reference-Format}
\bibliography{sample}

\end{document}